# Silver(I) complexes with nitrile ligands: new materials with versatile applications.

Karolina Gutmańska[a], Piotr Szweda[b], Marek Daszkiewicz[c], Tomasz Mazur[d], Konrad Szaciłowski[d], Anna Ciborska[a], Anna Dołęga[a*]

[a] Department of Inorganic Chemistry, Chemical Faculty,
Gdansk University of Technology, Narutowicza 11/12, 80-233 Gdansk, Poland *anndoleg@pg.edu.pl
[b] Department of Pharmaceutical Technology and Biochemistry, Gdańsk University of Technology, Narutowicza 11/12, 80-233 Gdańsk, Poland
[c] Institute of Low Temperature and Structure Research, Polish Academy of Sciences, Okólna St. 2, 50-422 Wrocław, Poland
[d] Academic Centre for Materials and Nanotechnology, AGH University of Science and Technology,
Department of Semiconductors Photophysics and Electrochemistry
Kawiory St. 30, 30-059 Krakow, Poland

**ABSTRACT**

In the present study, the structure, thermal stability, conductive properties, and antimicrobial activity of silver(I) complexes with nitrile ligands were investigated. For the construction of the materials, 2-cyanopyridine (2-cpy), 4-cyanopyridine (4-cpy), 1,2-dicyanobenzene (1,2-dcb), and 1,3-dicyanobenzene (1,3-dcb) were used in addition to the silver nitrite and nitrate. Four new compounds were isolated and structurally characterized: one molecular complex $[Ag_4(1,2\text{-}dcb)_4(NO_3)_4]$, two 1-D coordination polymers $[Ag_2(2\text{-}cpy)_2(NO_2)_2]_\infty$, $[Ag_3(1,3\text{-}dcb)_2(NO_3)_2]_\infty$ and one 3-D coordination polymer $[Ag(4\text{-}cpy)(NO_2)]_\infty$. The results indicate that the nitrite complexes display good antimicrobial properties against the tested bacterial and fungal strains. The presence of weakly coordinating CN groups increases the release of silver ions into the bacterial and yeast cell environments. Moreover, these materials exhibit unusual electrical properties in thin-layer devices. On the other hand, the nitrite and nitrate counterions give rise to the low thermal stability of the complexes.

## 1. Introduction

Elemental silver and its salts have been known for centuries as antibacterial agents.[1] A simple aqueous solution of $AgNO_3$ is still routinely used for the prophylaxis of neonatorum ophthalmia in infants[2,3], being effective in a concentration of 0.1%.[2] Silver sulfadiazine, a treatment of heavy burns, is considered an essential medicine by the World Health Organization.[4] Antimicrobial silver dressings prevent or treat infections in acute and chronic wounds.[5] Furthermore, while active against a spectrum of microorganisms, silver ions are relatively nontoxic to human cells.[1]

The construction of silver complexes with the best bactericidal, fungicidal, or anticancer properties comes down to the basic parameters considered at the initial stage of the synthetic effort. These include the solubility, stability in water, lipophilicity, redox reactivity, and the release rate of silver ions. These properties can be controlled by the selection of suitable ligands or by the use of biodegradable and biocompatible particle transport media. Several groups of organic molecules have been tested to form such compounds. As silver(I) is electron deficient, it readily combines with electron donating groups, commonly N-, O-, and P-donors, as well as carbene ligands.[6-15] The important group of ligands includes N-heterocyclic compounds, N-heterocyclic carbenes, phosphines, and amino acids.[6] Silver compounds with N-heterocyclic ligands are widely recognised as antimicrobial agents.[3,7-9] Contrary compounds containing thiol groups –SH neutralize the activity of silver nitrate against *Pseudomonas aeruginosa*.[7]

The increased antimicrobial activity of silver nitrile complexes resulting from the facile release of $Ag^+$ ions from labile complexes was suggested by Han *et al*.[16] So far, we have obtained and described the molecular structures of several molecular compounds and coordination polymers linked by cyanopyridine and dicyanobenzene ligands, and initially investigated their luminescent properties.[17] Within the present study, we are expanding the library of the nitrile complexes of silver with new molecular and polymeric species. Moreover, we study the antimicrobial activity of the large group of silver nitrile complexes and juxtapose the results with the antimicrobial activity of simple silver salts and organic nitriles.

The materials with relatively labile metal ions inspired us to investigate another property and application of these compounds. In addition to their antimicrobial properties, we describe preliminary research on the possibility of the formation of conductive filaments upon electrical stimulation of thin layers of nitrile silver complexes. The reversible change of the conductive properties may lead to applications very different from our initial idea, for example, thin-layer memristors.[18]

**2. Materials and Methods**

*2.1 General Information about Chemicals*

The following chemicals were used as purchased: silver(I) nitrate(V) $AgNO_3$, p.a., POCh; sodium nitrate(III) $NaNO_2$, p.a., B&K; 2-cyanopyridine (2-cpy); 4-cyanopyridine (4-cpy); 1,2-dicyanobenzene (1,2-dcb); 1,3-dicyanobenzene (1,3-dcb), acetonitrile, 99.9%, Merck; toluene, p., POCh; ethanol, 99.8%, POCh. $AgNO_2$ crystals were obtained as previously[19] and stored in a dark place. The flasks in which the reactions were carried out were protected from light by aluminum foil. However, we did not notice particular photo-sensitivity of the complexes, and there was no need to protect them against exposure to artificial light during standard laboratory operations such as filtering or weighing.

*2.2 Synthetic Procedures*

The cyanopyridine complexes crystallize after few days up to two weeks from the initial solutions (recipes below). The dicyanobenzene complexes crystallize after one to two months. The removal (evaporation) of acetonitrile (compounds **1**, **2**) or EtOH (**4**) from the initial solution accelerates the crystallization of the complexes.

**[Ag(4-cpy)(NO$_2$)]$_\infty$ (1)** Complex **1** was synthesized by the addition of 4-cyanopyridine (0.12 g, 1.2 mmol) in acetonitrile (4 mL) to the AgNO$_2$ solution (0.09 g, 5.8 mmol) in hot water (8 mL). The reaction mixture was stirred and left for crystallization. A white crystalline product was obtained in the form of blocks; yield 54%; mp. 144.9°C. Elemental analysis of C$_6$H$_4$AgN$_3$O$_2$: calcd. N 16.29, C 27.93, H 1.56; found N 16.12, C 27.86, H 1.56. FT-IR: 3109(w), 3091(w), 3064(w), 3044(w), 2250(m), 1599(s), 1553(w), 1494(w), 1417(m), 1330(m), 1288(m), 1240(vs), 1225(vs), 1213(vs), 1194(s), 1069(m), 1005(m), 962(w), 819(s), 759(w), 785(m), 558(s) cm$^{-1}$.

**[Ag$_2$(2-cpy)$_2$(NO$_2$)$_2$]$_\infty$ (2)** Complex **2** was synthesized in the same way as complex **1**, with the use of 2-cpy instead of 4-cpy, white crystalline product was obtained in the form of blocks/needles; yield 71%; mp. 106-107°C. Elemental analysis of C$_{12}$H$_8$AgN$_5$O$_2$: calcd. N 19.14, C 38.94, H 2.25; found N 19.14, C 38.94, H 2.26. FT-IR: 3440 (w), 3091(m), 3071(w), 3018(w), 2236(m), 1586(vs), 1570(w), 1466(s), 1431(s), 1272(vs), 1252(vs), 1207(m), 1153(w), 1093(m), 1051(m), 1003(s), 913(w), 828(w), 780(vs), 738(w), 637(w), 548(s) cm$^{-1}$.

**[Ag$_3$(1,3-dcb)$_2$(NO$_3$)$_2$]$_\infty$ (3)** Complex **3** was synthesized in the form of a mixture of products by the addition of solution 1,3-dicyanobenzene (0.068 g, 0.53 mmol) in toluene (7 mL) to AgNO$_3$ (0.09 g, 0.53 mmol) solution in ethanol (7 mL). The reaction mixture was stirred and left for crystallization. Crystals of **3** were separated under the optical microscope. Due to the difficulties in separating the pure complex, the product **3** was characterized exclusively by X-ray diffraction and FT-IR spectroscopy and the yield of the reaction was not determined. FT-IR: 3108(w), 3077(m), 3045(w), 2926(w), 2264(w), 2238(m), 1601(w), 1575(w), 1483(s), 1459(s), 1426(s), 1297(vs), 1179(w), 1021(w), 921(w), 905(w), 807(m), 673(m) cm$^{-1}$.

**[Ag$_4$(1,2-dcb)$_4$(NO$_3$)$_4$] (4)** Complex **4** was synthesized in the same way as complex **3** with the use of 1,2-dcb instead of 1,3-dcb. A colorless crystalline product was obtained in the form of blocks, yield 40%; mp. 128.5-129°C. Elemental analysis of C$_{16}$H$_8$Ag$_2$N$_6$O$_6$: calcd. N 14.10, C 32.24, H 1.35; found N 13.98, C 32.28, H 1.41. FT-IR (crystalline product): 3094(w), 3074(w), 3028(w), 2963(w), 2258(m), 2245(m), 1585(w), 1482(w), 1422(m), 1379(m), 1293(vs), 1261(s), 1228(w), 1211(m), 1168(m), 1089(w), 1034(m), 967(m), 817(m), 792(s), 776(s), 709(w) cm$^{-1}$.

For the synthesis of compounds: **[Ag(3-cpy)$_2$(NO$_2$)]**, **[Ag$_3$(3-cpy)$_2$(NO$_2$)$_3$]$_\infty$**, **[Ag(3-cpy)$_2$NO$_3$]$_\infty$**, **[Ag$_2$(1,4-dcb)(NO$_3$)$_2$]$_\infty$**, whose thermal, antimicrobial and electrical properties were studied within this work, please refer to Gutmańska and co-workers.[17]

*2.3 Physicochemical Methods*

FTIR spectra were recorded for the pure, crystalline products using a Nicolet iS50 equipped with Specac Quest diamond ATR device. All FTIR spectra were collected and formatted by OMNIC software. Elemental CHNS analyses were performed on a Vario EI Cube Elemental Analyser. The melting point of the compounds were determined by Stuart Scientific SMP3.

*2.4 Crystallography*

The crystal structure data of **1**, **3** and **4** were collected on an IPDS 2T dual beam diffractometer (STOE & Cie GmbH, Darmstadt, Germany) at 120.0(2) K with MoK$_\alpha$ radiation from a microfocus X-ray source (GeniX 3D Mo High Flux, Xenocs, Sassenage, France). Crystals were cooled using a Cryostream 800 open-flow nitrogen cryostat (Oxford Cryosystems). The crystallographic data are collected in Table 1.

*Table 1 Crystallographic data for compounds 1–4.*

| Complex | 1 | 2 | 3 | 4 |
|---|---|---|---|---|
| Formula | $C_6H_4AgN_3O_2$ | $C_{12}H_8AgN_5O_2$ | $C_{16}H_8Ag_2N_6O_6$ | $C_{16}H_8Ag_2N_6O_6$ |
| Formula weight | 257.99 | 362.10 | 596.02 | 596.02 |
| Temperature (K) | 120 | 295 | 120 | 120 |
| Wavelength (Å) | 0.71073 | 0.71073 | 0.71073 | 0.71073 |
| Crystal system | Orthorhombic | Orthorhombic, | Monoclinic | Monoclinic |
| Space group | $P2_12_12_1$ | $Pna21$ | $Ia$ | $P2_1/n$ |
| *a (Å)* | 6.5686(7) | 12.9728(2) | 14.721(4) | 6.9787 (14) |
| *b (Å)* | 9.2584(6) | 11.8412(2) | 3.7081(11) | 24.173 (3) |
| *c (Å)* | 12.6650(8) | 9.14655(16) | 33.908(7) | 11.1205(18) |
| *α (°)* | 90 | 90 | 90 | 90 |
| *β (°)* | 90 | 90 | 93.130(18) | 91.562 (15) |
| *γ (°)* | 90 | 90 | 90 | 90 |
| *V (Å)* | 770.22(11) | 1405.04(4) | 1848.2(8) | 1875.3 (5) |
| *Z* | 4 | 2 | 4 | 4 |
| Crystal size (mm) | 0.23× 0.20 × 0.13 | 0.237 × 0.133 × 0.125 | 0.32 × 0.22 × 0.07 | 0.14 × 0.11 × 0.09 |
| $T_{min}$, $T_{max}$ | 0.594, 0.740 | 0.723, 0.846 | 0.575, 0.841 | 0.773, 0.845 |
| $\mu$ (mm$^{-1}$) | 2.57 | 1.442 | 2.17 | 2.14 |
| Absorption correction | Integration | Gaussian | Integration | Integration |
| Reflections collected/unique/unique[I > 2r(σ)] | 7408, 1500, 1484 | 37767, 3210, 2581 | 11298, 3731, 2242 | 16658, 3685, 3308 |
| $R_{int}$ | 0.021 | 0.027 | 0.072 | 0.019 |
| Data/restraints/parameters | 1500/0/111 | 3210/1/181 | 3731/2/272 | 3685/0/271 |
| Goodness of fit (GOOF) on F² | 1.027 | 1.070 | 0.994 | 1.016 |
| Final R indices [I > 2r(σ)] | R1 = 0.0235<br>wR2 = 0.0615 | R1 = 0.0422<br>wR2 = 0.0656 | R1 = 0.0538<br>wR2 = 0.1273 | R1 = 0.0179<br>wR2 = 0.0426 |
| R indices (all data) | R1 = 0.0237<br>wR2 = 0.0616 | R1 = 0.0299<br>wR2 = 0.0721 | R1 = 0.0962<br>wR2 = 0.1538 | R1 = 0.0219<br>wR2 = 0.0439 |
| Δρmax, Δρmin (e Å$^{-3}$) | 0.52 / − 0.50 | 0.526 / − 0.589 | 1.091 / − 0.969 | 0.32 / − 0.38 |
| CCDC numbers | 2254449 | 2254450 | 2254451 | 2254452 |

Data collection and image processing for **1**, **3**, **4** were performed with X-Area 1.75.[20] Intensity data were scaled with LANA (part of X-Area) to minimize differences of intensities of symmetry equivalent reflections (integration method). Structures were solved using intrinsic phasing procedure implemented in SHELXT and all nonhydrogen atoms were refined with anisotropic displacement parameters by the full matrix least squares procedure based on $F^2$ using the SHELX–2014 program package.[21] The Olex[22] and Wingx[23] program suites were used to prepare the final version of CIF files.

The X-ray diffraction data of **2** were collected at room temperature on Oxford Diffraction four-circle single-crystal diffractometer equipped with a CCD detector using graphite-monochromatized MoK$_\alpha$ radiation The raw data were treated with the CrysAlis Data Reduction Program (version 1.171.39.46). The intensities of the reflections were corrected for Lorentz and polarization effects. Absorption correction was applied taking into account the unit cell content and optimizing the crystal shape.

Hydrogen atoms in **1**-**4** were refined using isotropic models with $U_{iso}(H) = 1.2\ U_{eq}(C)$. Olex[22] and Mercury[24] were used to prepare all the figures.

Of the measured monocrystals three were of excellent quality (merging error $R_{int}$<3%) and one, namely compound **3**, exhibited $R_{int}$=7.2%, which is an accepted value. The suggested model of **3** converged to R1 = 5.8%. This was the crystal that was selected from the mixture of products under the optical microscope and we were unable to prepare crystals of better quality (see Chapter 2.2 Synthetic Procedures).

*2.5 Thermal Stabilities*

The thermal analysis was performed using an SDT 650 thermoanalyser from TA Instruments. Thermograms were recorded in a synthetic air atmosphere with a heating rate of 10 °C/min to 1000°C. The mass of the samples used in the analyses was within 8–40 mg. The thermoanalytical curves were analyzed using the Origin computational program (version 9.0, OriginPro).

*2.6 Antimicrobial Tests*

Silver(I) complexes and ligands were tested for their antimicrobial activity against five reference strains of bacteria, including two strains of Gram positive staphylococci, namely *S. aureus* ATCC 25923 and *S. aureus* ATCC 29213 and three strains of pathogenic Gram negative bacteria, namely *P. aeruginosa* ATCC 27853, *E. coli* ATCC 25922, *S. enterica* PCM 2266. Furthermore, the antimicrobial potential of synthesized agents was evaluated against two reference strains of yeast pathogens *C. albicans* SC5314 and *C. glabrata* DSM II 226.

Two different assays were applied: determination of minimum inhibitory concentration (MIC) of substances of interest towards the bacterial strains growing in suspension and determination of growth inhibition zones on agar medium (agar disc-diffusion method) were applied for evaluation of antibacterial and antifungal activity of synthesized compounds.

MIC values were determined in 96-well titration plates by the two-fold broth microdilution method according to the CLSI standard methodologies.[25,26] For all compounds the activity was evaluated in the concentration range of 256.0 to 0.5 µg/mL The assay was performed using Mueller Hinton Broth (MHB) (Sigma-Aldrich) for bacteria and RPMI 1640 medium supplemented with 2% glucose and buffered to pH 7.0 with MOPS buffer (3-*N*-morpholinopropanesulfonic acid) for yeast. Plates were incubated 24 h under static conditions at 37°C. The growth intensity (optical density of the medium in the wells) of the bacteria/yeast was measured using a Victor3 Plate reader (Perkin Elmer, Waltham, MA, USA). The lowest concentration of the compound that caused at least 90% growth inhibition of bacteria/yeast (compared to growth observed in MHB/RPMI medium not supplemented with any compound) was taken as the MIC value.

In the other assays, Petri plates (ϕ=90mm) with Mueller Hinton Agar 2 (for bacteria) or RPMI medium solidified with 2% agar (for yeast) were inoculated by reference strains of bacteria/yeast. Inoculation was performed by streaking with a sterile cotton swab soaked in a suspension of each microorganism reference strain (final optical density of each suspension $OD_{600}$ = 0.1) freshly prepared in sterile PBS (phosphate buffered saline, pH 7.4). Subsequently, up to six paper discs (ϕ=5mm) soaked in 20 µl of solution of the silver complexes (10.24 mg/mL) were placed on the surface of the inoculated agar medium. The plates were incubated over night at 37 °C and zones of inhibition of bacteria/yeast around the discs were observed and measured.

## 2.7 Electric Measurements

The representatives of the family of compounds were spin-coated (SPIN 150i, Polos) on an unpatterned conductive substrate (ITO glass slide, Ossila Ltd). Investigated materials were compounds **1**, **2** and **4**. Additionally, one of the complexes explored previously was measured: **([Ag$_3$(3-cpy)$_2$(NO$_2$)$_3$]$_\infty$**.[17] The average thickness of the films were 56 ± 8 nm. In the next preparation step, 100 nm thick Ag electrodes of dimensions (1.3mm x 1.5mm) were sputtered (Leica EM ACE600) at a top of the complex layer. In each case, the I-V curves characteristic measurements were conducted under an ambient atmosphere. Current–voltage characteristic (I–V) were performed on a Biologic SP-300 system with two electrodes connected to the ITO substrate and Ag top electrodes, respectively. For the I–V measurements, the DC voltage sweep was executed ranging the voltage limits alongside the scan velocities; see the following section.

## 2.8. Quantum Chemical Modelling

DFT calculations were performed using the Gaussian 16 Revision C.01 software package using the B3LYP functional and TZVP basis set[27,28] in vacuum. Unconstrained optimisation of the molecular geometry was carried out under tight convergence criteria ($\Delta E_{SCF} \leq 10^{-8}$ hartree). Potential distribution maps were plotted using the GaussView software package, version 5.0.8.

## 3. Results and Discussion

### 3.1 Syntheses

As a result of the syntheses, we have obtained three coordination polymers **1**, **2**, **3** and one molecular complex **4** (Scheme I).

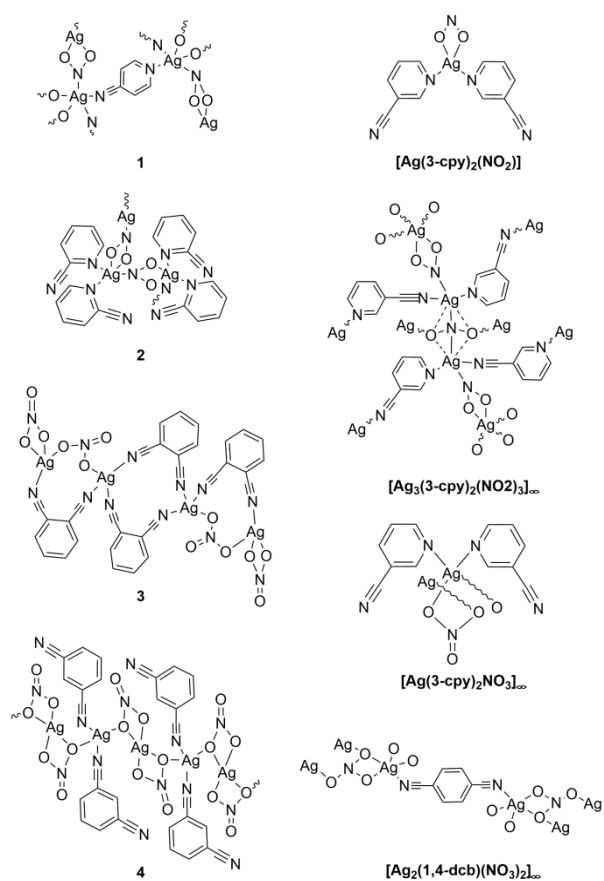

**Scheme I.** *Formulas of coordination polymers **1, 2, 3** and molecular complex **4** synthesized within this work (left side); formulas of silver complexes with 3-cyanopyridine and 1,4-dicyanobenzene synthesized previously (right side,[17]).*

The syntheses of the complexes were simple and required mixing of the reagents: AgNO$_2$ or AgNO$_3$ and nitrile in the appropriate solvents in the molar ratio 1:1. In the case of compound **3**, we did not find the conditions appropriate for crystallization of the pure product. Using different solvents and molar ratios of the reactants, we usually obtained a 1,3-dcb ligand as the major product of crystallization. Finally, the crystals of complex **3** were separated under the optical microscope and a suitable single-crystal for X-ray diffraction experiment was found among them. In Scheme I we also show the formulas of the complexes obtained previously [17], whose selected properties were studied and described in this paper.

### *3.2 Crystal Structures*

Compound **1** crystallizes in orthorhombic symmetry, space group *P*2$_1$2$_1$2$_1$ with one 2-cyanopyridine molecule and one nitrite ion in the asymmetric part of the unit cell (Figure 1a). The coordination number of silver is formally equal to 5 and the geometry is close to a square pyramid, as indicated by τ$_5$=0.01.[29] Three-dimensional pattern in the crystal structure of **1**, can be described as antiparallel chains connected by nitrite ions (along *b* axis) or 4-cyanopyridine molecules (along *c* axis). The separation of silver atoms *via* nitrite is 5.218 Å and the distance between silver atoms linked by 4-cyanopyridine is 9.983 Å. The interchain Ag---Ag distances between the antiparallel chains are 6.659 Å. The packing and the chains within it are illustrated in Figure 1b.

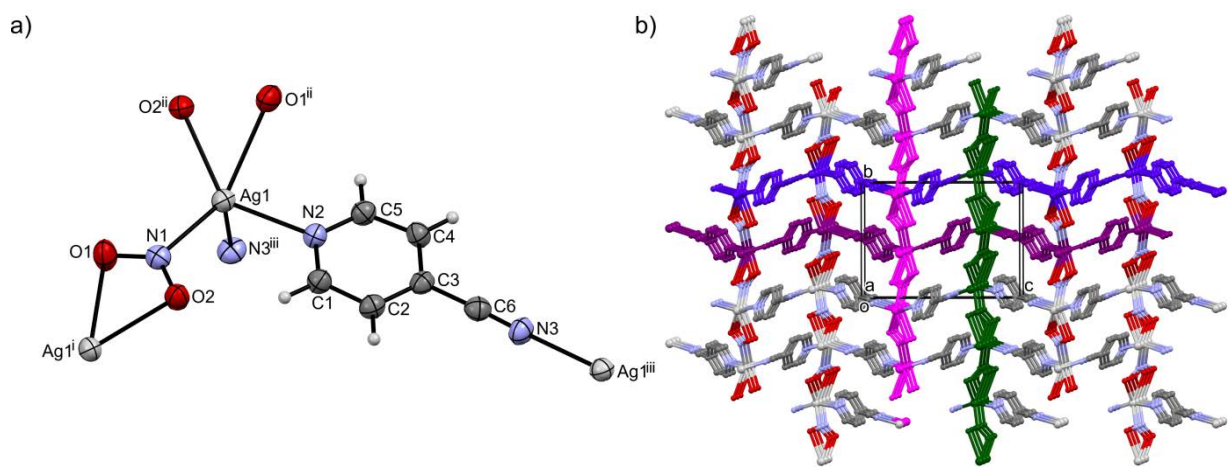

*Figure 1. Molecular structure of complex **1**: a) coordination sphere of the Ag(I) ion with the atomic numbering scheme, displacement ellipsoids drawn at the 50% probability level. Important bond lengths [Å]: Ag1–N1 2.337(4), Ag1–N2 2.359(4), Ag1–O1$^{ii}$ 2.595(4), Ag1–O2$^{ii}$ 2.384(3), Ag1–N3$^{iii}$ 2.362(4). Important angles [°]: N1–Ag1–N2 103.02(13), N1–Ag1–O1$^{ii}$ 135.31(13), N1–Ag1–O2$^{ii}$ 108.49(14), N1–Ag1–N3$^{iii}$ 103.46(15), N2–Ag1–O1$^{ii}$ 85.69(12), N2–Ag1–O2$^{ii}$ 135.62(13), N2–Ag1–N3$^{iii}$ 90.82(15), O1$^{ii}$–Ag1–O2$^{ii}$ 49.94(11), O1$^{ii}$–Ag1–N3$^{iii}$ 120.36(15), O2$^{ii}$–Ag1–N3$^{iii}$ 110.8(1); symmetry operations: $^{i}$ =2-x, -1/2+y, 1.5-z, $^{ii}$ =2-x, 1/2+y, 1.5-z, $^{iii}$ =1/2-x, 1-y, ½+z; b) packing of polymeric structure, hydrogen atoms of pyridine ring removed for clarity; the antiparallel chains of silver atoms linked via nitrite ions indicated with magenta and green and these linked by 4-cyanopyridine drawn in purple and violet-blue.*

In complex **2**, the silver atom is coordinated by two molecules of 2-cyanopiridine and two bridging nitrite ions that form a distorted tetrahedral environment: τ$_4$ = 0.85, τ$_4$' = 0.82 (Figure 2a).[29,30] The nitrite ion bridges two silver ions *via* a short Ag1–N5 bond of 2.262(7) Å, and long one Ag1–O1 2.66(1) Å. Silver atoms and nitrite ions form a skeleton of the 1-D coordination polymer that is propagated along the crystallographic axis *c*. The polymer exhibits a zigzag configuration as shown in Figure 2b. The shortest Ag1⋯Ag1 distance in the chain is 5.289(2) Å. The polymeric structure is stabilized by the π-π stacking interactions of the aromatic rings of the 2-cyanopyridine ligands of the neighboring chains (Figure 3).

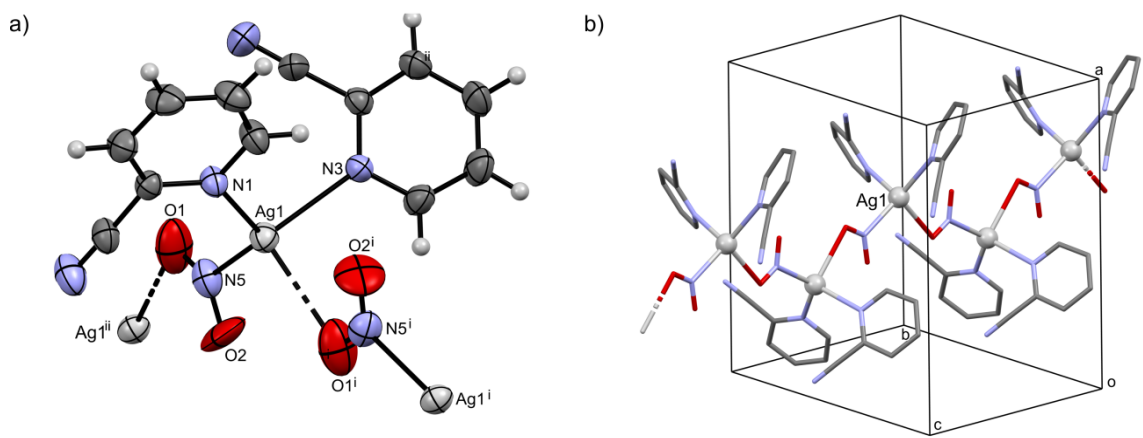

*Figure 2. Molecular structure of complex 2: a) coordination sphere of the Ag(I) ion with the atomic numbering scheme, displacement ellipsoids drawn at the 50% probability level. Important bond lengths [Å]: Ag1–N1 2.327(5), Ag1–N3 2.435(4), Ag1–N5 2.262(7), Ag1–O1$^i$ 2.66(1). Important angles [°]: N1–Ag1–N3 108.64(15), N1–Ag1–N5 125.1(2), N3–Ag1–N5 107.60(19), N1–Ag1–O1$^i$ 115.01(30), N3–Ag1–O2$^i$ 96.98(20), N5–Ag1–O2$^i$ 99.69(30); symmetry operations: $^i$ =1-x, 1-y, -1/2+z, $^{ii}$ =1-x, 1-y, 1/2+z; b) 1-D polymeric structure, hydrogen atoms omitted for clarity.*

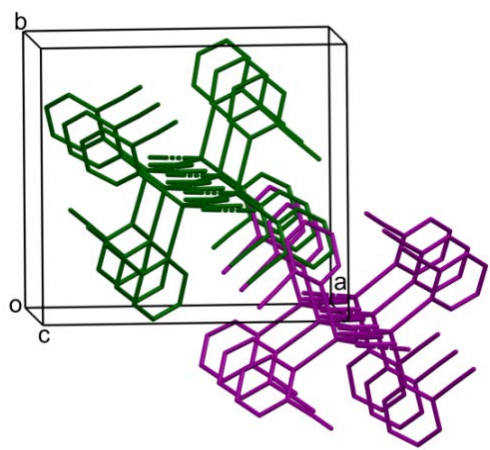

*Figure 3. Crystal packing of complex 2; hydrogen atoms of pyridine ring removed for clarity, cell axes demonstrated; the parallel chains of silver atoms linked via nitrite ions indicated with magenta and green.*

Compound **3** has the one-dimensional polymeric structure of alternating cationic [Ag(1,3-dcb)$_2$]$^+$ units and [Ag(NO$_3$)$_2$]$^−$ anions. In the asymmetric part of the unit cell, two silver(I) ions are present with a distorted tetrahedral geometry (Figure 4a). The Ag1 atom is coordinated by two molecules of 1,3-dcb and two oxygen atoms from nitrate ions ($\tau_4$ = 0.74, $\tau_4'$ = 0.63), while the Ag2 atom is coordinated exclusively by the oxygen atoms of two nitrate ions ($\tau_4$ =0.72, $\tau_4'$ =0.69). Distorted coordination is affected by chelation of the nitrate anion in the asymmetric mode, where the longer Ag2–O5 bond is 2.560 Å long and the shorter one Ag2–O1 is equal to 2.300 Å. The Ag–O–N–Ag chain propagates along the *a* axis and exhibits a zigzag configuration characteristic for 1,4-dcb complexes of 11 group metals.[31] The shortest Ag···Ag separation within the chain is 4.840 Å and between the chains it is 3.708 Å, which, being longer than the sum of the van der Waals radii of silver atoms (3.44 Å[32]), excludes argentophilic interactions.[33] The arrangement of the polymer chains within the crystal is illustrated in Figure 4b.

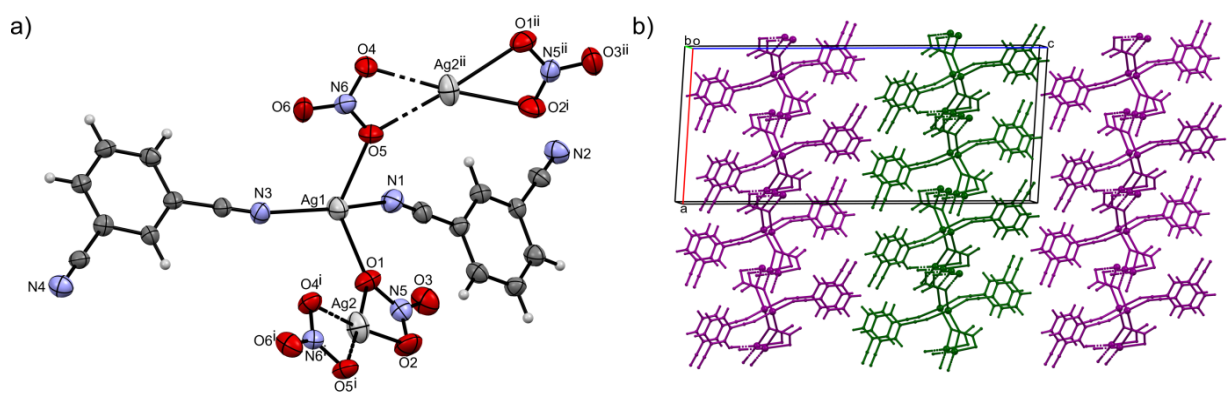

*Figure 4. Molecular structure of complex 3: a) coordination spheres for two symmetry independent Ag(I) ions. Displacement ellipsoids drawn at the 30% probability level. Important bond lenghts [Å]: Ag1–O1 2.558(15), Ag1–O5 2.519(14), Ag1–N1 2.198(16), Ag1–N3 2.208(15), Ag2–O1 2.574(17), Ag2–O2 2.398(18), Ag2–O4 2.418(12), Ag2–O5 2.567(14). Important angles: O1–Ag1–O5 109.1(4), O1–Ag1–N1 102.1(6), O1–Ag1–N3 99.9(5), O5–Ag1–N1 87.5(6), O5–Ag1–N3 108.8(5), N1–Ag1–N3 146.3(5), O1–Ag2–O2 51.0(5), O1–Ag2–O4 125.7(4), O1–Ag2–O5 176.3(5), O2–Ag2–O4 175.2(7), O2–Ag2–O5 132.6(5), O4–Ag2–O5 50.7(4); symmetry operations: i = 1/2+x, -2-y, z, ii = -1/2+x, -2-y, z; b) crystal packing viewed along b axis.*

Compound **4** crystallizes in the monoclinic crystal system, space group *P*2$_1$/*n*. It is a molecular, tetranuclear complex, in which four silver cations are bridged by four 1,2-dcb ligands (Figure 5a). The molecule lies on the inversion centre and therefore the asymmetric part of the unit cell contains two silver cations, two nitrate anions and two 1,2-dcb molecules. Both independent Ag1 and Ag2 ions feature C.N. = 4, with distorted geometries. Atom Ag1 is coordinated by one –C≡N group of 1,2-dcb and three oxygen atoms of two nitrates (one chelating). The geometrical indices of Ag1 are $\tau_4$ = 0.61, $\tau_4$′ = 0.57. Atom Ag2 is coordinated by three –C≡N groups and one nitrate oxygen and its geometry is closer to tetrahedral ($\tau_4$ = 0.77, $\tau_4$′ = 0.69). Ag–N bonds are in the range 2.178 Å to 2.333 Å, while Ag-O bonds fall within the range 2.431 – 2.496 Å. The important bond lengths and angles are given in Fig. 5 caption. "Wavy" molecules of **4** pack in piles parallel to axis *a* (Figure 5b). The intermolecular aggregation is stabilized by weak C–H···O hydrogen bonds between 1,2-dcb and nitrate – almost all C–H bonds of 1,2-dcb are engaged in relatively short H-bonds of this type (*e.g.* C5–H5···O1 3.193$_{C5···O1}$/2.567$_{H5···O1}$ Å). The interactions are reinforced by several Ag1···π contacts (*e.g.* Ag1···C6 3.225 Å) between the neighboring molecules. There are no close Ag···Ag contacts within or between the molecules.

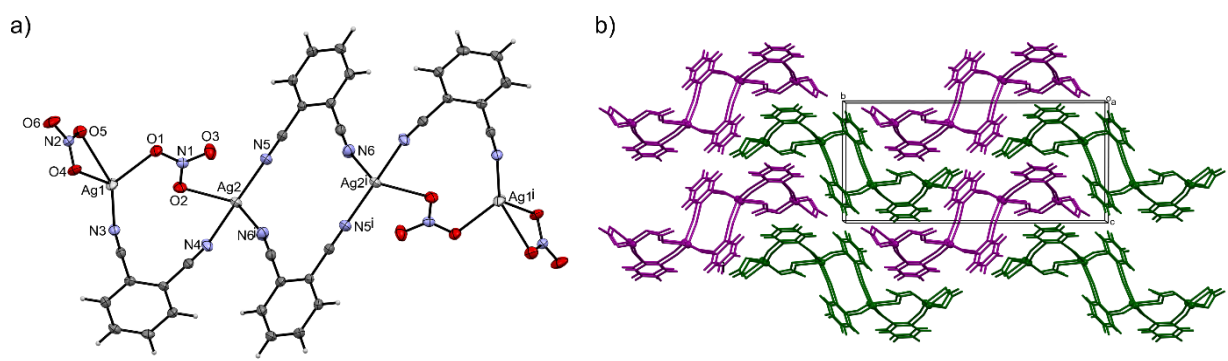

***Figure 5*** *Molecular structure of complex **4**: a) molecule with the numbering scheme, displacement ellipsoids drawn at the 50% probability level; important bond lenghts [Å]: Ag1–O1 2.4307(16), Ag1–O4 2.4262(15), Ag1–O5 2.4956(16), Ag1–N3 2.1783(19), Ag2–O2 2.4423(16), Ag2–N5 2.2116(19), Ag2–N6$^i$ 2.3330(19); Symmetry operations: $^i$ = 2-x, 1-y, 2-z; b) crystal packing viewed along a axis.*

### 3.3 FTIR Spectroscopy

Infrared (IR) spectroscopy was used to confirm the coordination behaviour of the nitrile ligands in the bulk samples. FTIR ATR spectra of complexes **1**–**4** compared to their respective ligands are presented in Figs. 1S – 4S of Supplementary Information. In the absence of coordination to Ag$^+$ ions, we should expect a similar value of the stretching vibration of the nitrile group of the coordinated and free ligand; in the case of cyanopyridines, this means that they coordinate only through the N atom of the pyridine ring.[17,34] Direct coordination of the nitrile group to the metal ion shifts the maximum of the band associated with the $v_{CN}$ mode in comparison to the uncoordinated nitriles.[17,35] Analysis of the FT-IR spectra of the ligands and their silver complexes **1**-**4** revealed that the differences in the nitrile stretching frequencies for cyanopyridines and their complexes are in accordance with the crystal structures. The characteristic frequencies are collected in Table 2.

***Table 2*** *The frequencies of the $v_{C≡N}$ stretching mode of free and complexed cyanopyridines and cyanobenzenes and complexes **1**–**4**.*

| Compound | $v_{CN}$ [cm$^{-1}$] | $\Delta v_{C≡N}$ [cm$^{-1}$] vs. non-coordinated ligand | Comment |
|---|---|---|---|
| **4-cpy** | 2235 | – | |
| **1** | 2250 | +15 | Coordinated nitrile |
| **2-cpy** | 2243 | – | |
| **2** | 2236 | -7 | Non-coordinated nitrile |
| **1,3-dcb** | 2234 | – | |
| **3** | 2264 | +30 | Coordinated nitrile |
| | 2238 | +4 | Non-coordinated nitrile |
| **1,2-dcb** | 2232 | – | |
| **4** | 2258 | +26 | Coordinated nitrile |
| | 2245 | +13 | Coordinated nitrile |

FTIR spectra of complexes **1** and **2** (cyanopyridine ligands) showed the presence of a coordinated -C≡N group in complex **1** and a noncoordinated nitrile in the case of complex **2**. The FTIR spectrum of dicyanobenzene complex **3** features two distinct bands of nitrile stretching indicating the presence of coordinated (+30 cm$^{-1}$) and non-coordinated (+4 cm$^{-1}$) nitrile. In complex **4**, both nitrile groups are coordinated to silver ions, and in the spectrum we observe two bands at +13 and +30 cm$^{-1}$ compared to the FTIR spectrum of the free ligand.

The bands due to $v_a$ and $v_s$ frequencies of the nitrite ion in **1** and **2** are seen at about 1365-1332 and 1264-1235 cm$^{-1}$ respectively and the bending frequency around 847-845 cm$^{-1}$.[36-38] The asymmetric ($v_3$, several bands centered at 1322 cm$^{-1}$) and symmetric ($v_1$, around 1040 cm$^{-1}$) modes of nitrates **3** and **4** split into several bands, especially in complex **4**, which features both bridging and terminal nitrate ions (Figures 5, 3S, and 4S).

*3.4 Thermal Stabilities*

The thermal stability was studied for the **1**-**2**, **4** and silver complexes described previously: **[Ag(3-cpy)$_2$(NO$_2$)]**, **[Ag$_3$(3-cpy)$_2$(NO$_2$)$_3$]$_\infty$**, **[Ag(3-cpy)$_2$NO$_3$]$_\infty$**, **[Ag$_2$(1,4-dcb)(NO$_3$)$_2$]$_\infty$** [17] in air and N$_2$ atmosphere. The TGA and DTA data for all compounds are presented in Figs. 5S – 22S and summarized in Tables 1S and 2S of Supporting Information.

Analyzing the stability of compounds derived from substrates such as AgNO$_3$ containing cyanopyridine and dicyanobenzene molecules, we should expect endothermic effects to accompany the decomposition of complexes. Previous studies have shown that AgNO$_3$ itself can decompose *via* different pathways, which can lead to the formation of diverse products.[39,40] The main, confirmed, one-step decomposition pathway leads to metallic silver, nitrogen oxides NO$_2$, NO, metallic silver nanoparticles and molecular oxygen.[40] The other products: N$_2$ and N$_2$O, silver oxide/silver nitrate complex suggested by Paulik and coworkers[39] were not confirmed later by Otto.[40] The DTA curve of AgNO$_3$ shows two endothermic peaks at approximately 165 and 210°C. The latter peak was attributed to the melting of the compound and the lower peak is probably connected with a polymorphic transition.[39-41] In the case of the TG and DTG curves, complete decomposition of the liquified compound is observed between 360-500°C, with the final endothermic peak at 500°C and the formation of Ag, regardless of the atmosphere applied in the experiment.[40] The thermal analysis of AgNO$_2$ shows that decomposition process starts at 120°C/128°C and initially results in silver(I) oxide and nitrogen oxides NO and NO$_2$.[42,43] At this elevated temperature the released nitrogen oxides react quickly with the silver(I) oxide producing silver nitrate.[42]

On the other hand, for metal complexes with cyanopyridines, TG and DTA curves were characterized by three endothermic peaks attributed to the partial release of cyanopyridine molecules from the compound structure.[44,45] In the case of complexes containing dicyanobenzene molecules, TG/DTA diagrams with three peaks were also obtained, however, the loss of dcb molecule was associated with only one of the decomposition peaks between 118 and 175°C.[46,47]

The complete TG, DTG, and DTA data for the studied complexes are placed in Supplementary Materials. The TG and DTG curves revealed a very complex way of thermal decomposition of the studied complexes. The obtained DTA data showed that some or even all of the decomposition processes are exothermic reactions, which differs significantly from the data in the literature. These exothermic steps were observed both in the air and in the protective atmosphere of nitrogen, thus the oxidation by the oxygen was excluded.[48,49] However, it is very probable that the exothermic effects are due to the reactions between the released oxidizing nitrogen oxides and the organic nitriles. Furthermore, silver ions may have a catalytic effect on reactions as the transiently formed silver(I) oxide is a long-recognized oxidation agent.[50,51]

Compound **1** decomposes in three steps (Figs. 5S and 6S). Decomposition begins above 110°C with the release of nitrogen oxides and partial release of 4-cpy. It is represented by a sharp exothermic gradient in the DTA curve with a maximum at 153°C. The second stage of decomposition occurs at 226-327°C and is accompanied by the loss of 4-cpy. The third stage, which led to the final decomposition of complex **1**, occurred at temperatures above 327°C. The slight loss in mass is probably associated with the release of oxygen from the resulting $Ag_2O$ molecule. It is evident from the DTA curve that the two final stages were exothermic – either almost unnoticeably ($T_{max}$ not defined) or strongly ($T_{max\ DTA}$=344°C).

A similar decomposition pattern was observed for the complex **2** and the silver nitrite coordination polymer **[Ag$_3$(3-cpy)$_2$(NO$_2$)$_3$]$_\infty$**, where the first stage of the DTG curve was characterized by a sharp gradient and two smaller peaks for the other stages (Figs. 7S-10S). The decomposition of complex **2**, starts above 71°C with a maximum DTA of 122°C, the second process undergoes at 152-290°C ($T_{max\ DTA}$=260°C), and the third stage begins at 299°C ($T_{max\ DTA}$=391°C). The weight loss of the first stage corresponds to the evolution of nitrogen oxides, mainly NO, and the release of one molecule of 2-cpy. In later stages, a partial release of 2-cpy is observed. The decomposition of **[Ag$_3$(3-cpy)$_2$(NO$_2$)$_3$]$_\infty$**, begins above 80°C ($T_{max\ DTA}$=155°C) with the detachment of the nitrite group and the 3-cpy fragment (Figs. 9S and 10S). The subsequent steps are characterized by a small mass loss associated with further decomposition of the organic ligand. The second step corresponds to the temperature range 190-287 ($T_{max\ DTA}$=286°C), and the third undergoes within 293-350 °C ($T_{max\ DTA}$=320°C).

Finally, in terms of the thermal stability of silver nitrite complexes, we have analyzed the thermal decomposition process of the molecular silver nitrite, cyanopyridine complex **[Ag(3-cpy)$_2$(NO$_2$)]**. Contrary to the coordination polymers described above, molecular **[Ag(3-cpy)$_2$(NO$_2$)]** featured two endothermic processes at lower temperatures, which may correspond to phase transitions such as melting and evaporation (Figs. 11S and 12S). At higher temperatures again exothermic reactions are observed with $T_{max\ DTA}$ at 152°C and 336°C. Compared to coordination polymers of similar composition it is apparent that the molecular compound is less thermally stable, *i.e.* the changes begin at lower temperature. Additionally, when we repeated the measurement in the protective atmosphere of nitrogen, the degradation was less complex and the exothermic effects were less pronounced, thus we conclude that in the elevated temperature the complex reacts with oxygen (Figs. 13S and 14S).

The decomposition of nitrate complex **4** in the atmosphere of air, illustrated in Figs. 15S and 16S, begins at 131°C and this first stage ends at 249°C ($T_{max\,DTA}$=247°C); further weight loss occurs between 250-310°C ($T_{max\,DTA}$=306°C). On the DTA diagram, at least one endothermic and two exothermic peaks are clearly visible. The endothermic peak at 134°C (DTA) roughly corresponds to the melting point determined by the usual method (128.5-129°C). Each step of weight loss is accompanied by the exothermic peak on the DTA at 247 and 306°C. Interestingly, the results of the TG/DTG analysis of **4** in $N_2$ atmosphere were very similar (Figs. 17S and 18S), which confirms that the exothermic reactions undergo within the decomposing complex without the participation of oxygen.

For comparison, we performed and describe the TG and TA analysis of two more nitrate silver coordination polymers with nitrile ligands: **[Ag(3-cpy)$_2$NO$_3$]$_∞$** and **[Ag$_2$(1,4-dcb)(NO$_3$)$_2$]$_∞$**.[17] Both compounds decomposed in a similar manner but differently from molecular **4**. The three distinct steps of weight loss were accompanied by two endothermic and one exothermic peaks on the DTA chart (Figs. 19S-22S).

In the end, we indicate one more feature that differentiates the behavior of the described silver complexes during thermal analysis. In most cases the mass percentage of the residue after thermal decomposition roughly corresponds to the metallic silver as in nitrites: **2**, **[Ag$_3$(3-cpy)$_2$(NO$_2$)$_3$]$_∞$**, **[Ag(3-cpy)$_2$(NO$_2$)]** or silver oxide as in nitrates: **[Ag(3-cpy)$_2$NO$_3$]$_∞$** and **[Ag$_2$(1,4-dcb)(NO$_3$)$_2$]$_∞$**. There are two compounds that definitely do not follow this pattern: complex **1** in which the residue is 30% instead of 42% expected for metallic silver and compound **4**, which is volatile since the solid residue is only 6,4% instead of calculated 26% (Table 1S).

### *3.5 Antibacterial and Antifungal Activity*

Antimicrobial tests were performed for complexes **1**, **2**, **4** as well as **[Ag(3-cpy)$_2$(NO$_2$)]**, **[Ag$_3$(3-cpy)$_2$(NO$_2$)$_3$]$_∞$**, **[Ag(3-cpy)$_2$NO$_3$]$_∞$**, **[Ag$_2$(1,4-dcb)(NO$_3$)$_2$]$_∞$**.[17] All compounds demonstrated relatively good antifungal and antibacterial activity with the MIC values in the 1-64 μg/mL range (Table 3). The coordination polymer **[Ag$_3$(3-cpy)$_2$(NO$_2$)$_3$]$_∞$** exhibited the highest activity - it effectively inhibited *P. aeruginosa* at the concentration of 8 μg/mL which is 3 times lower than MIC value of AgNO$_2$. Satisfactory antibacterial properties were also displayed by complex **1** and the molecular complex **[Ag(3-cpy)$_2$(NO$_2$)]**. Both complexes inhibited growth of *P. aeruginosa* at a concentration 16 μg/mL which is 2 times lower than AgNO$_2$. Complexes **2**, **4** and **[Ag(3-cpy)$_2$NO$_3$]$_∞$** exhibited the same or even lower activity than AgNO$_2$ and AgNO$_3$. Other strains of bacteria exhibited slightly higher resistance to the activity of the complexes studied. With the exception of *S. enteritica* and complex

**[Ag$_3$(3-cpy)$_2$(NO$_2$)$_3$]$_\infty$** (MIC=128 µg/mL), the MIC values of all compounds were in the range of 16 to the 64 µg/mL, comparable to MIC values of AgNO$_2$ and AgNO$_3$.

Taking into account the MIC values, the highest activity was observed against two fungal pathogens, *C. albicans* and *C. glabrata*. Almost all compounds effectively inhibited the growth of fungal strains at concentrations of 1-4 µg/mL. Only the molecular complex **[Ag(3-cpy)$_2$(NO$_2$)]** was less active (MIC 16 or 32 µg/mL). However, it should be emphasized that *C. albicans* and C. *glabrata* exhibited a higher susceptibility to AgNO$_2$ and AgNO$_3$ (MIC in the range of 1-2 µg/mL).

*Table 3 Results of the antibacterial activity of Ag(I) compounds – MIC [µg/ml]*

| Ag(I)complex | Bacteria | | | | | Fungi | |
| --- | --- | --- | --- | --- | --- | --- | --- |
| | Gram (+) | | Gram (-) | | | | |
| | S. aureus ATCC 25923 | S. aureus ATCC 29213 | P. aeruginosa ATCC 27853 | E. coli ATCC 25922 | S. enterica PCM 2266 | C. albicans SC 5314 | C. glabrata DSM II 226 |
| **1** | 32 | 16 | 16 | 32 | 32 | 2 | 2 |
| **2** | 32 | 16 | 32 | 32 | 32 | 2 | 2 |
| **4** | 32 | 32 | 32 | 32 | 64 | 16 | 32 |
| **[Ag(3-cpy)$_2$(NO$_2$)]** | 64 | 64 | 16 | 64 | 128 | 2 | 4 |
| **[Ag$_3$(3-cpy)$_2$(NO$_2$)$_3$]$_\infty$** | 32 | 32 | 8 | 32 | 64 | 1 | 2 |
| **[Ag(3-cpy)$_2$NO$_3$]$_\infty$** | 64 | 64 | 64 | 64 | 32 | 2 | 4 |
| **[{Ag(1,2-dcb)$_4$}(NO$_3$)$_4$]** | 32 | 32 | 32 | 16 | 32 | 2 | 2 |
| **AgNO$_2$** | 32 | 16 | 32 | 32 | 16 | 2 | 2 |
| **AgNO$_3$** | 16 | 32 | 32 | 8 | 16 | 1 | 1 |
| **3-cpy/2-cpy/ 4-cpy/ 1,2-dcb/1,4-dcb** | >256 | >256 | >256 | >256 | >256 | >256 | >256 |

*Table 4 Results of the antibacterial activity of Ag(I) compounds – diameter of the inhibition zone [mm].*

| Ag(I)complex | Bacteria | | | | | Fungi | |
| --- | --- | --- | --- | --- | --- | --- | --- |
| | Gram (+) | | Gram (-) | | | | |
| | S. aureus ATCC 25923 | S. aureus ATCC 29213 | P. aeruginosa ATCC 27853 | E. coli ATCC 25922 | S. enterica PCM 2266 | C. albicans SC 5314 | C. glabrata DSM II 226 |
| **1** | 13 | 8 | 14 | 8 | 20 | 20 | 24 |
| **2** | 8 | 7 | 13 | 7.1 | 20 | 16 | 18 |
| **4** | 12 | 10 | 14 | 9.5 | 24 | 15 | 18 |
| **[Ag(3-cpy)$_2$(NO$_2$)]** | 7.5 | 7.5 | 14 | 9.5 | 24 | 24 | 18 |
| **[Ag$_3$(3-cpy)$_2$(NO$_2$)$_3$]$_\infty$** | 14 | 8 | 15.5 | 10 | 24 | 20 | 17 |
| **[Ag(3-cpy)$_2$NO$_3$]$_\infty$** | 7 | 7 | 11 | 10 | 24 | 21 | 17 |
| **[Ag$_2$(1,4-dcb)(NO$_3$)$_2$]$_\infty$** | 8 | 7 | 14 | 9 | 23 | 14 | 17 |
| **AgNO$_2$** | 8 | 7 | 13 | 10 | 22 | 14 | 14 |
| **AgNO$_3$** | 10 | 7 | 13.5 | 10 | 23 | 16 | 16 |
| **3-cpy/2-cpy/ 4-cpy/ 1,2-dcb/1,4-dcb** | 0 | 0 | 0 | 0 | 0 | 0 | 0 |

Interestingly, in a different type of assay, *i.e.* disc diffusion, the studied complexes revealed selective high activity against *S. enteritica* with diameter zones in the range of 20-24 mm, which is an importantly higher value compared to other investigated strains of bacteria (Table 4). Another interesting result of this part of the study is the increased antifungal activity of nitrile complexes compared to AgNO$_2$ and AgNO$_3$. The remaining results of the agar disc diffusion method are in agreement with the MIC dilution assay, for example, high activity against *P. aeruginosa*, particularly in the case of complex **[Ag$_3$(3-cpy)$_2$(NO$_2$)$_3$]$_\infty$** (Tables 3 and 4).

### *3.6. Electrical characterization of thin films*

With regard to potential application of the synthesized compounds, outside antibacterial functionality scope, all of the measured nitrile silver complexes showed nonlinear electrical behaviour in a form of narrow, pinched I-V hysteresis loops. Typically in literature these are called hysteresis curves, and are associated to memristive, resistive switching behaviour. As can be concluded from Figure 6, in order to register a nonlinear response, particular potential threshold must be exceeded. In the case of **1** for smaller window (Figure 6a), only a linear response exists, while the nonlinear hysteretic response can be measured for the potential window spanning from <-6 V; +6V> (Figure 6b). In other words – the sample behaves as a typical resistor within a narrower potential window, whereas the extension of this window brings about hysteretic behaviour. This indicated the redox-type behaviour of the metal/complex interface. Despite minimal fluctuation of the switching points, in each individual measurement two distinguishable states of the devices can be found, either it is low resistivity state (LRS, higher current for defined voltage) or high resistivity state (HRS).

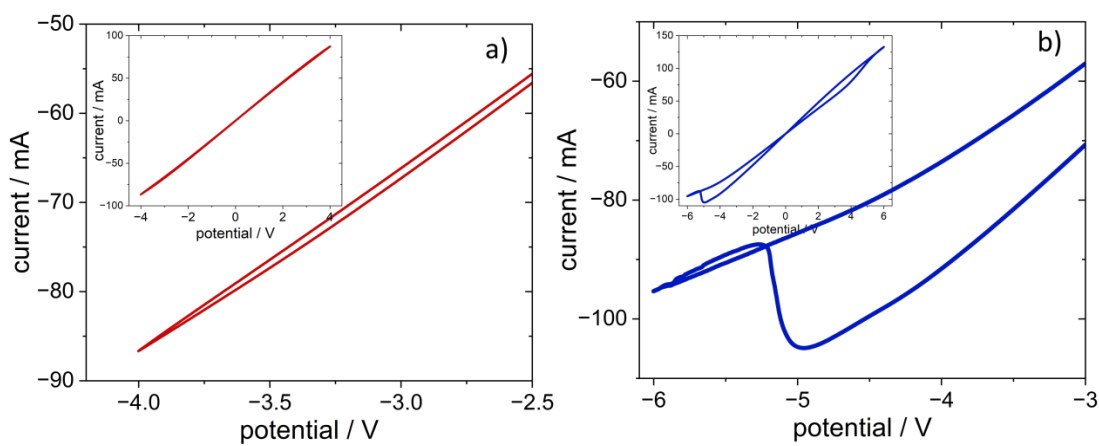

*Figure 6. Electric characterization from the thin film samples of complex **1**: a) range +/-4V and b) range +/- 6V. Each figure depicts results taken with 100 mV/s velocity scans after several cycles of pre-conditioning with lower voltages than showed. The insets demonstrate results of the full-range scans.*

From an electrical point of view, the most interesting compound to measure happens to be complex **1**. Hysteresis is clearly visible (Figure 6b) with additional capacitance or fractional memristance signals occurring upon measurement in quadrant III.[52,53] These results indicate that at least resistive switching processes[54] is responsible for resistive switching. However, it cannot be the formation of a metallic filament (electrode material). In such a case, the ratio (on/off ratio) between resistivity in HRS and resistivity in LRS would reach 10$^4$ or even more This effect disqualifies studied materials for memory application, where high ON/OFF ration is required. Fortunately, even low on/off efficient cases of thin film materials may serve information processing purposes.[55]

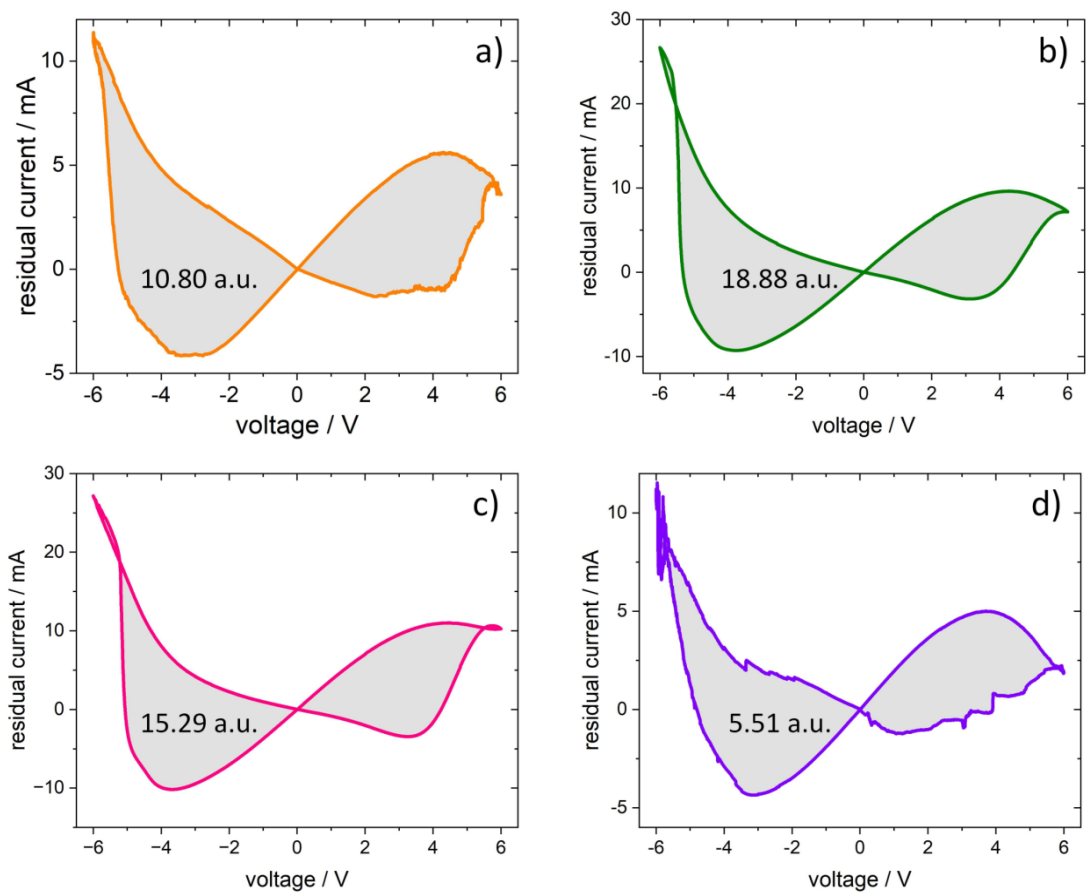

*Figure 7.* Conductivity hysteresis presented as a residual current for silver complexes with cyanopyridines and 1,2-dcb: a) complex **2** (with 2-cpy); b) **[Ag$_3$(3-cpy)$_2$(NO$_2$)$_3$]$_∞$**; c) complex **1** (with 4-cpy); and d) complex **4** (with **1,**2-dcb). The numbers indicate the total surface area of the hysteresis loop.

These rather narrow hysteresis loops can be better visualized and analyzed upon subtraction of the linear component (Ohmic component), which was approximated by a homogeneous linear function. The residuals of these adjustments are presented in Figure 7. Thus, the obtained hysteresis curves indicate memristive character of all devices: a pinched hysteresis loop, with significantly asymmetric lobes. This asymmetry, along with the differences of maximum and minimum currents (cf. Fig. 6) suggests a leaky Schottky barrier character of obtained junctions. The asymmetry factors (current ratio for forward and reversed bias voltages) are low (1.13, 1,36 and 1,42 for 2-, 3-, and 4-cyanopyridine complexes, respectively) but indicate weak current rectification. For 1,2-dicyanobenzene complex **4** this effect is even smaller (1.11). The memristive character of these junctions can be quantified by the total surface area within the hysteresis loop. The highest value (18.86 a.u.) was recorded for the 3-cyanopyridine complex **[Ag$_3$(3-cpy)$_2$(NO$_2$)$_3$]$_\infty$**, whereas 1,2-dicyanobenzene complex **4** had the lowest value (5.51 a.u.). Interestingly, hysteresis curves recorded for 3-cyanopyridine and 4-cyano-pyridine complexes show multiple cross points, indicating the contribution of capacitive effects or ferroelectric ordering during voltage scans. Taking into account that one crossing point is always present at U = 0 V, the second possibility seems to be more justified.[56,57] Furthermore, some theoretical studies relate these multiple cross points to memductance.[58]

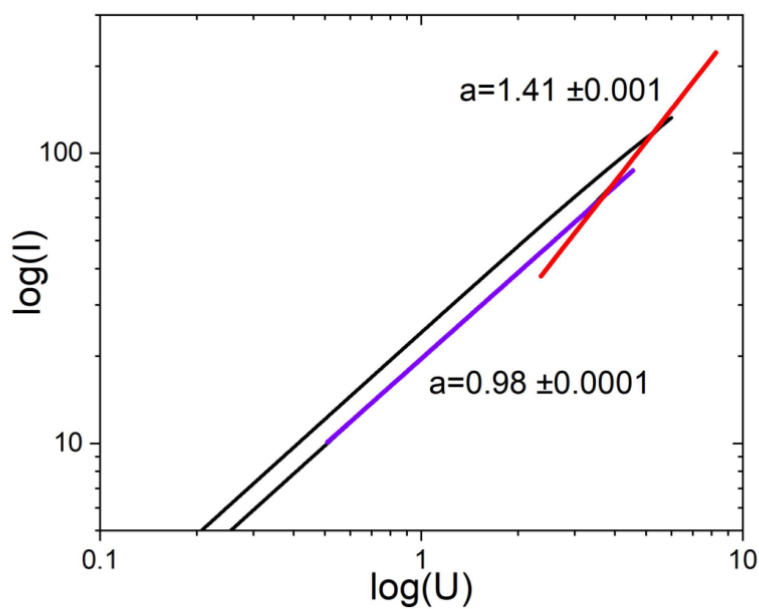

*Figure 8. Log-log graph of current-voltage characteristics for 4-cyanopyridine complex 2.*

The similarity in the slope values of these fits, along with almost the same thickness of the layer (56 ± 8 nm) indicates very similar specific conductivity of the compounds studied. It is justified by a similar composition. Log-log plots (Figure 8) indicate mostly Ohmic character of electron transport in these materials (current-voltage slope in log scale equal to unity), however at voltages higher than 4V, significant contribution from trap-assisted space-charge limited conductivity (SCLC) can be postulated on the basis of significantly higher slopes (1.17, 1.24 and 1.41 for 2-, 3-, and 4-cyanopyridine complexes, respectively). In the case of 1,2-dicyanobenzene, the unity slope was observed throughout the voltage range. Slopes significantly higher than unity indicate quadratic dependence of current versus voltage, characteristic for space-charge limited current transport mechanism, therefore they are layer called SCLC exponents.

The similarities in electrical properties of all cyanopyridine derivatives originate from similarities in ligand structures and properties, as well as some analogies in binding motifs. The most significant difference in all the studied ligands is their charge distributions and the resulting dipole moments (Figure 9). In all cases, the nitrile groups bear a significant negative charge, the same is also observed for heteroatoms. Various arrangements of substituents with respect to heteroatoms result in substantial changes in dipole moment and inhomogeneity of electric potential within the crystals. These inhomogeneities, in turn, influence charge transport and charge trapping phenomena at the metal/layer interface.[59]

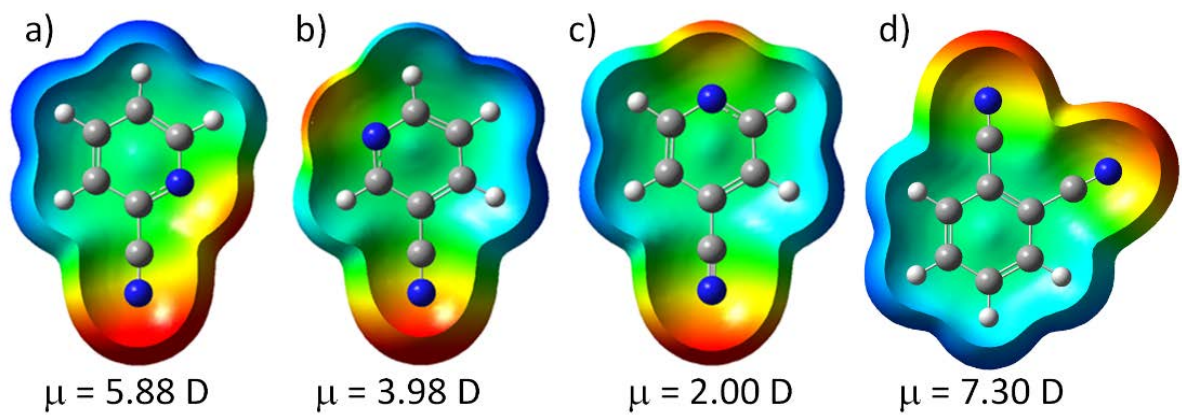

*Figure 9.* Electrostatic potential distribution maps projected over total electron density for: a) 2-cyanopyridine; b), 3-cyanopyridine; c), 4-cyanopyridine; d) and 1,2-dicyanobenzene. The numbers indicate the dipole moment of these molecules calculated at the B3LYP/TZVP level.

It can be noted that all physical quantities related to electrical properties of silver complex layers correlate with dipole moments of the ligands: the surface area of hysteresis loops decreases with increasing dipole moment; the same concerns the SCLC exponent (cf. Figure 8) and the rectification properties of Schottky junctions (Figure 10).

At the current stage of research it is difficult to unambiguously assign this dependence to one particular process, but putatively, correlation with dipole moments indicates the role of electric fields variations at molecular level, along the conductivity path, as well as the height of the Schottky barrier at the silver/material interface.

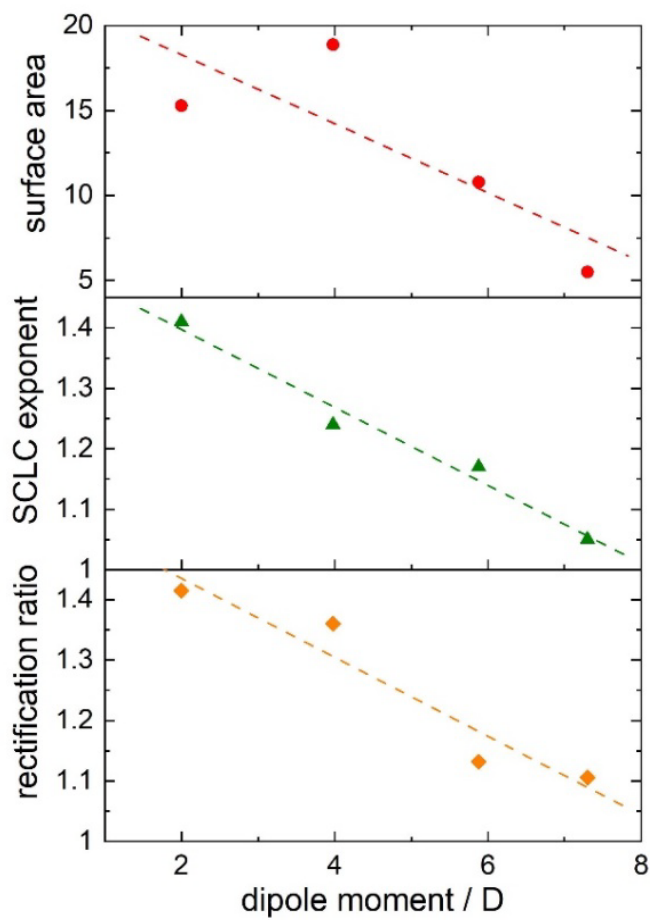

*Figure 10.* *Correlation of some electrical properties of Ag/complex/ITO junctions with dipole moments of the ligands.*

**4. Summary and Conclusions**

In this work, we have described four new silver complexes with nitrite/nitrate anions and two types of nitrile ligands: cyanopyridines and cyanobenzenes. We have studied the thermal properties of these new and other similar silver complexes to conclude that these compounds are not thermally stable and show a complicated pattern of decomposition, probably involving oxidation reactions between nitrite/nitrate ions and nitrile ligands. On the other hand, the studied nitrile complexes of silver show very good antimicrobial and antifungal properties, as anticipated by the weak binding and facile release of silver ions from the nitrile complex. Along with antimicrobial activity, interesting electrical phenomena have been observed that are related to the electronic structure of organic ligands. All of this makes the interest in the silver/nitrile systems worthwhile and encourages us to modify them to achieve materials with better stability and similar, valuable properties.

**5. Acknowledgments:**

The research was supported by the "Excellence Initiative - Research University" programs at Gdansk ´ University of Technology: RADIUM LEARNING THROUGH RESEARCH DEC-11/RADIUM/2021 (syntheses) and SILICIUM SUPPORTING CORE R&D FACILITIES DEC-2/2021/ IDUB/V.6/Si (crystallographic measurements). KG and AD have been supported by the National Science Centre (Poland) within the OPUS project (grant agreement No. UMO-2022/47/B/ST4/00728). TM and KS have been supported by the National Science Centre (Poland) within the OPUS project (grant agreement No. UMO-2020/37/B/ST5/00663). MD would like to thank ILT&SR PAS for financial support by statutory activity subsidy, no. 2019/5.## 6. Conflicts of Interest

The authors declare that they have no known competing financial interests or personal relationships that could have appeared to influence the work reported in this paper.

## 7. Data Availability Statement

Crystallographic data for the structure of complexes **1-4** in this paper have been deposited with the Cambridge Crystallographic Data Centre (CCDC), 12 Union Road, Cambridge CB21EZ, UK. Copies of the data can be obtained free of charge on quoting the depository numbers **2254449**-**2254452** for the complexes (see Table 1); Fax: +44-1223-336-033; mail: **deposit@ccdc.cam.ac.uk**, website **http://www.ccdc.cam.ac.uk**. FTIR and TG/DTG/DTA data are available in the **supporting information**. The remaining data that support the results of this work are accessible upon reasonable request from the corresponding author.

## 8. References

[1] S. Medici, M. Peana, V. M. Nurchi, M. A. Zoroddu, *J. Med. Chem.* **2019**, 62, 5923.

[2] H. Nishida, H. M. Risemberg, *Pediatrics*, **1975**, 56, 368.

[3] R. Rowan, T. Tallon, A. M. Sheahan, R. Curran, M. McCann, K. Kavanagh, M. Devereux, V. McKee, *Polyhedron* **2006**, 25, 1771.

[4] J. Kuchar, J. Rust, C. W. Lehmann, F. Mohr, *Inorg. Chem.* **2020**, 59, 10557.

[5] D. Leaper, *Int. Wound J.* **2012**, 9, 461.

[6] S. Medici, M. Peana, G. Crisponi, V. M. Nurchi, J. I. Lachowicz, M. Remelli, M. A. Zoroddu, *Coord. Chem. Rev.* **2016**, 327–328, 349.

[7] S. Y. Liau, D. C. Read, W. J. Pugh, J. R. Furr, A. D. Russell, *Lett. Appl. Microbiol.* **1997**, 25, 279.

[8] P. O. Asekunowo, R. A. Haque, M. R. Razali, S. W. Avicor, M. F. F. Wajidi, *Eur. J. Med. Chem.* **2018**, 150, 601.

[9] M. O. Karataş, N. Özdemir, M. Sarıman, S. Günal, E. Ulukaya, İ. Özdemir, *Dalton Trans.* **2021**, 50, 11596.21

# Silver(I) complexes with nitrile ligands: new materials with versatile applications.

Karolina Gutmańska[a], Piotr Szweda[b], Marek Daszkiewicz[c], Konrad Szaciłowski[d], Tomasz Mazur[d], Anna Ciborska[a], Anna Dołęga[a]

[a]Department of Inorganic Chemistry, Chemical Faculty,
Gdansk University of Technology, Narutowicza 11/12, 80-233 Gdansk, Poland *anndoleg@pg.edu.pl
[b] Department of Pharmaceutical Technology and Biochemistry, Gdańsk University of Technology, Narutowicza 11/12, 80-233 Gdańsk, Poland
[c] Institute of Low Temperature and Structure Research, Polish Academy of Sciences, Okólna St. 2, 50-422 Wrocław, Poland
[d] Academic Centre for Materials and Nanotechnology, AGH University of Science and Technology, Krakow, Poland

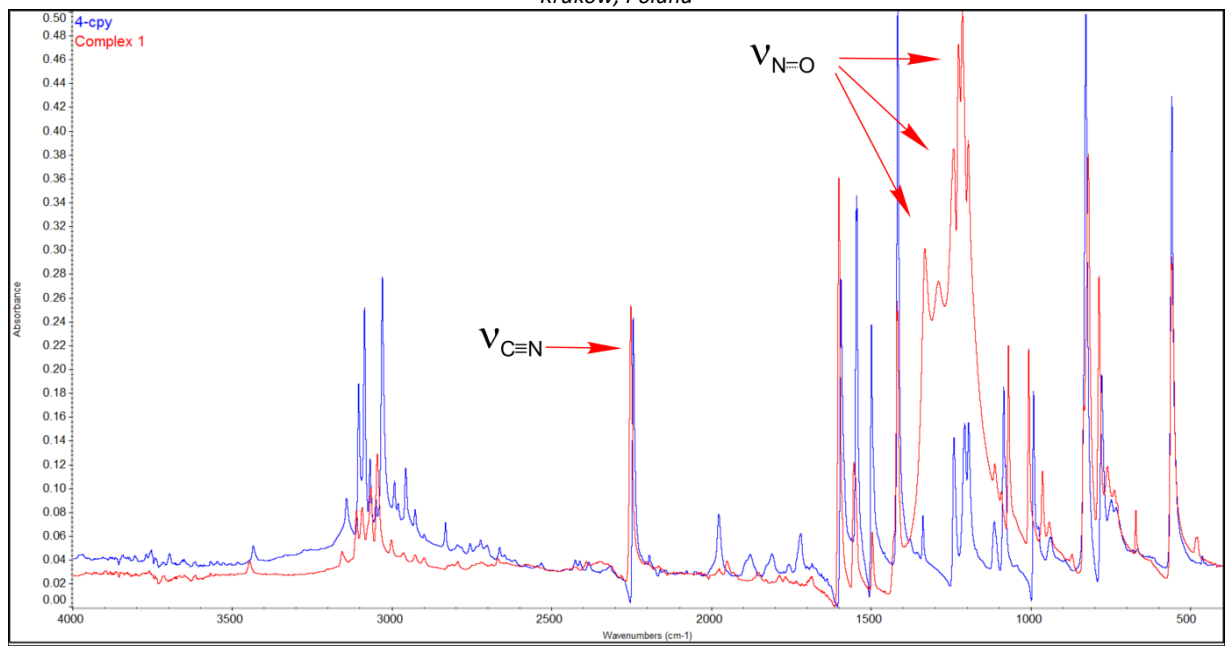

Fig.1S Ft-IR spectrum of Complex **1** and the corresponding ligand.

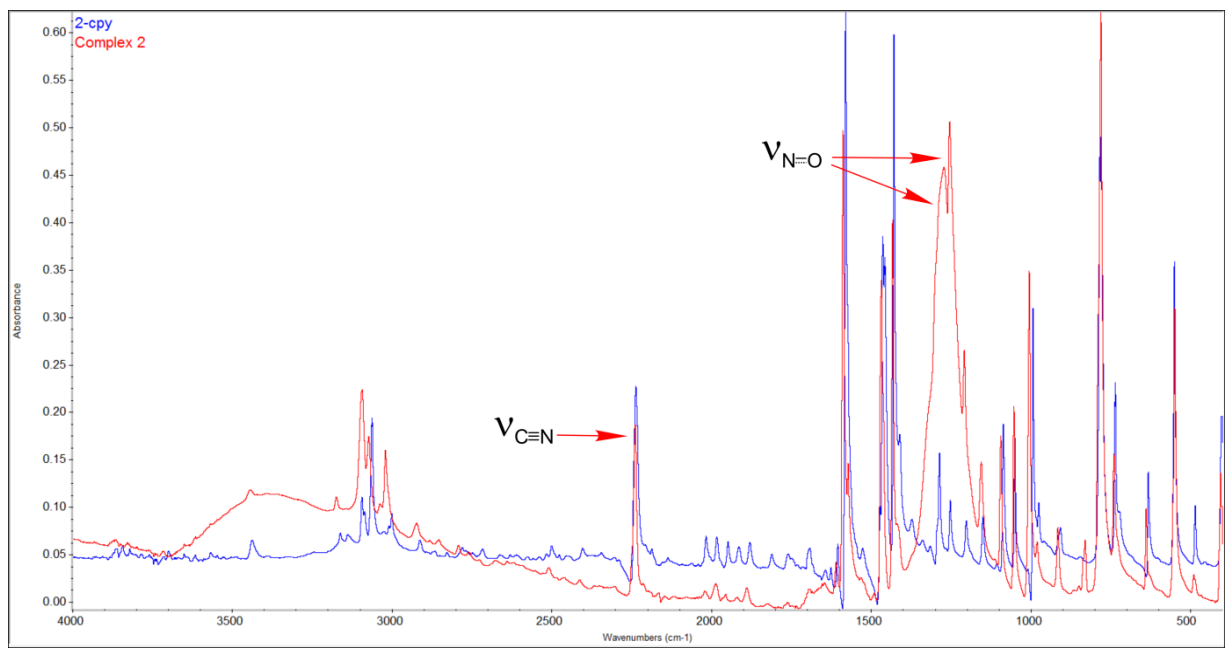

Fig.2S Ft-IR spectrum of Complex **2** and the corresponding ligand.

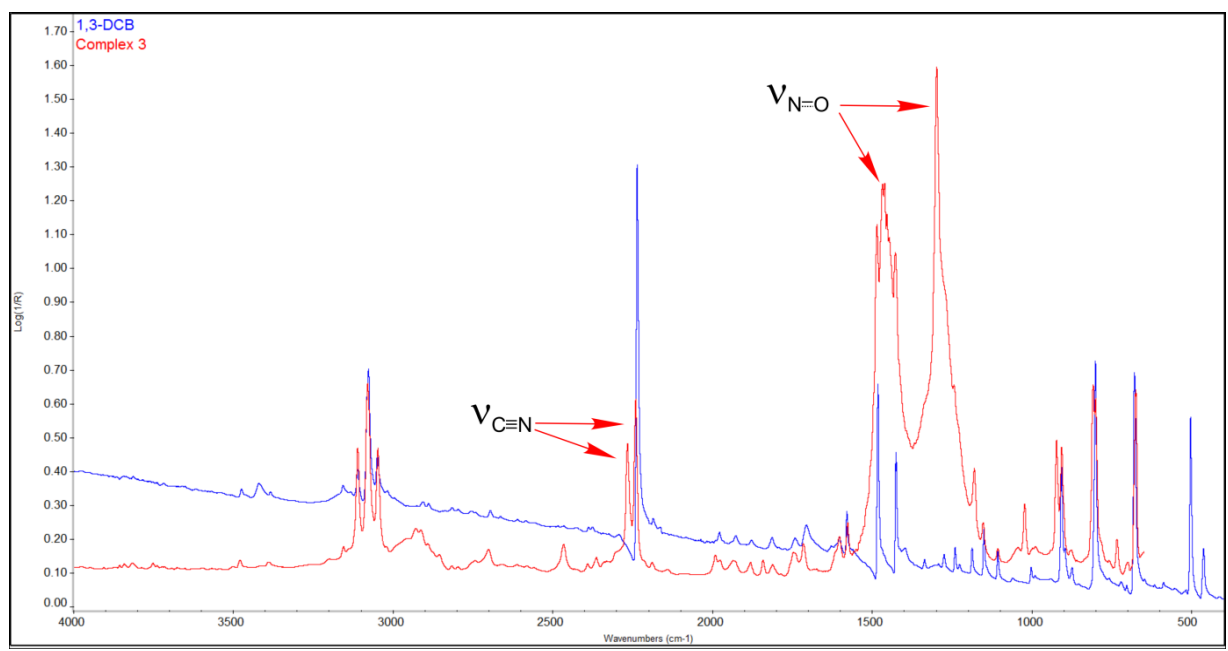

Fig.3S Ft-IR spectrum of Complex **3** and the corresponding ligand.

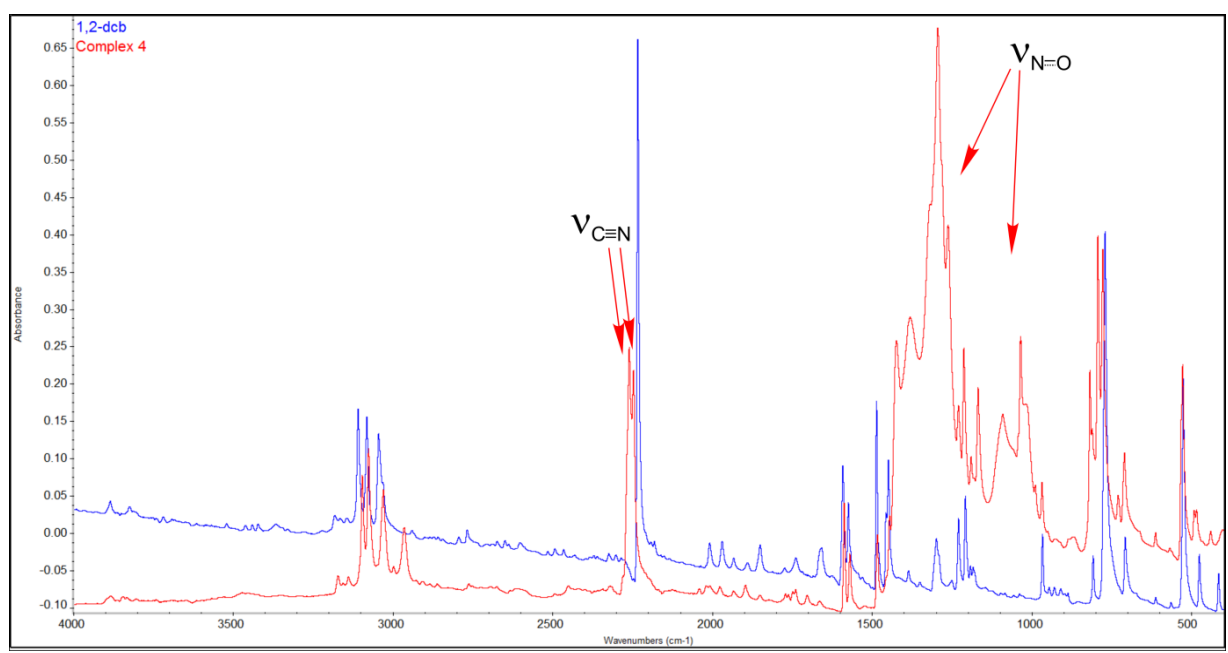

Fig.4S Ft-IR spectrum of Complex **4** and the corresponding ligand.

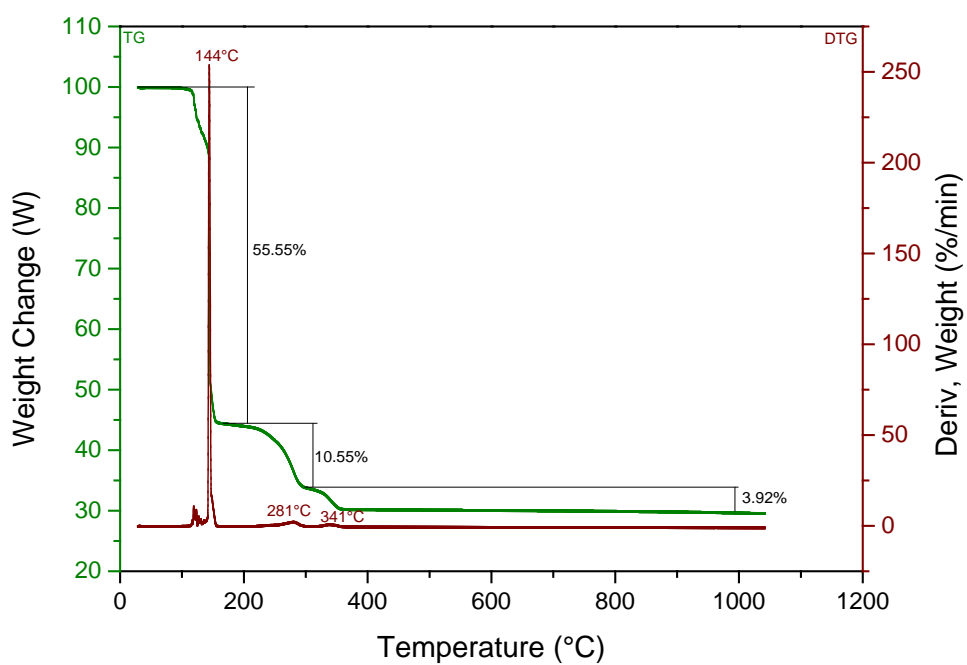

Fig. 5S TG and DTG derivative thermograms of **1** in air atmosphere.

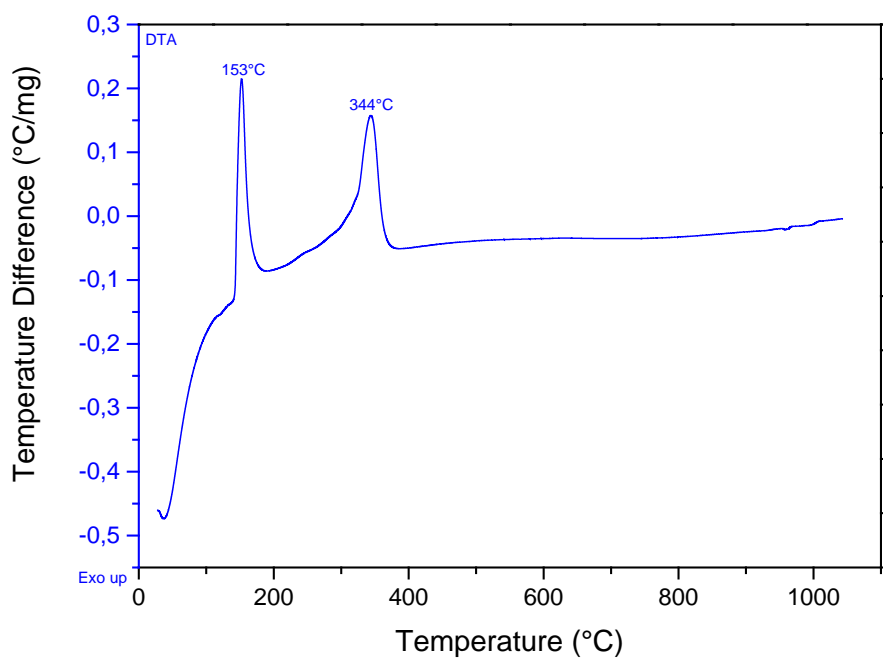

Fig. 6S DTA derivative thermograms of **1** in air atmosphere.

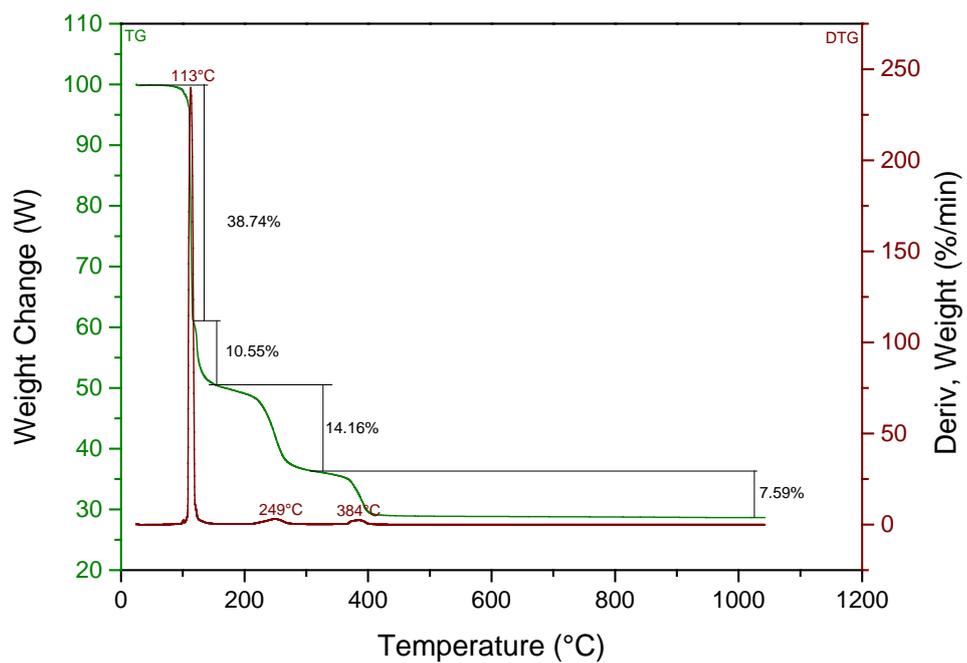

Fig. 7S TG and DTG derivative thermograms of **2** in air atmosphere.

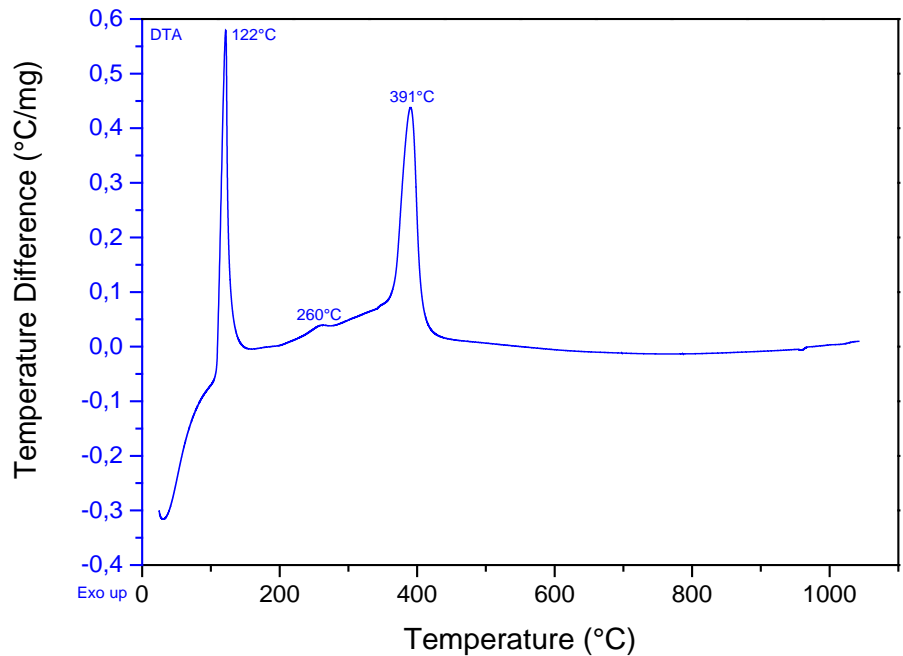

Fig. 8S DTA derivative thermograms of **2** in air atmosphere.

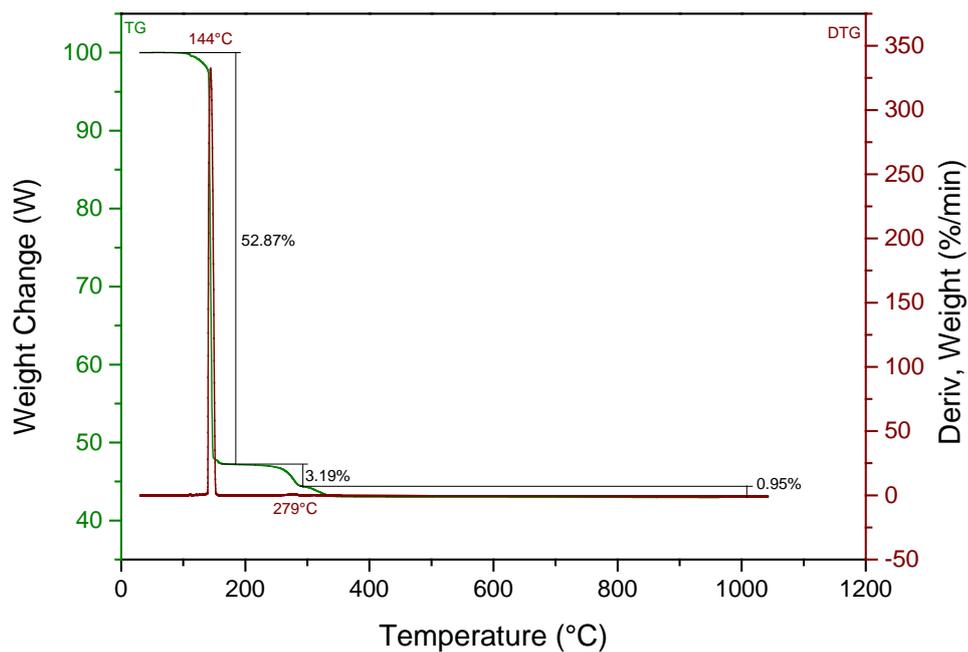

Fig. 9S TG and DTG derivative thermograms of **[Ag₃(3-cpy)₂(NO₂)₃]∞** in air atmosphere.

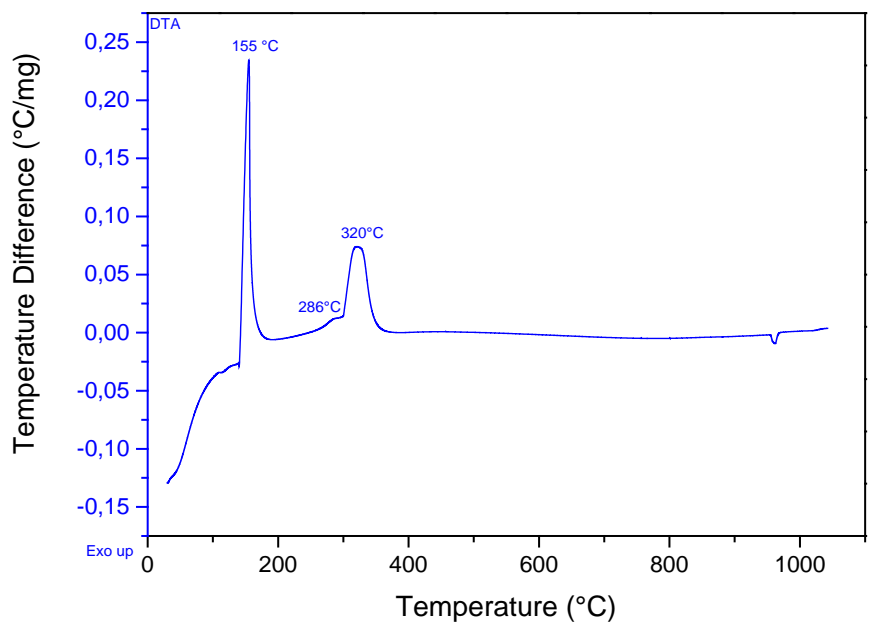

Fig. 10S DTA derivative thermograms of **[Ag₃(3-cpy)₂(NO₂)₃]∞** in air atmosphere.

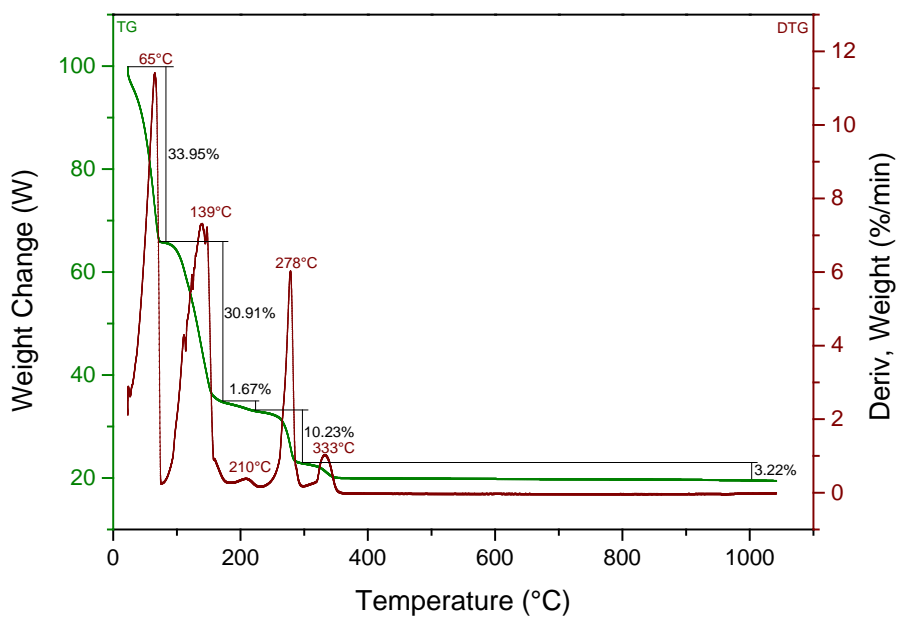

Fig. 11S TG and DTG derivative thermograms of **[Ag(3-cpy)₂(NO₂)]** in air atmosphere.

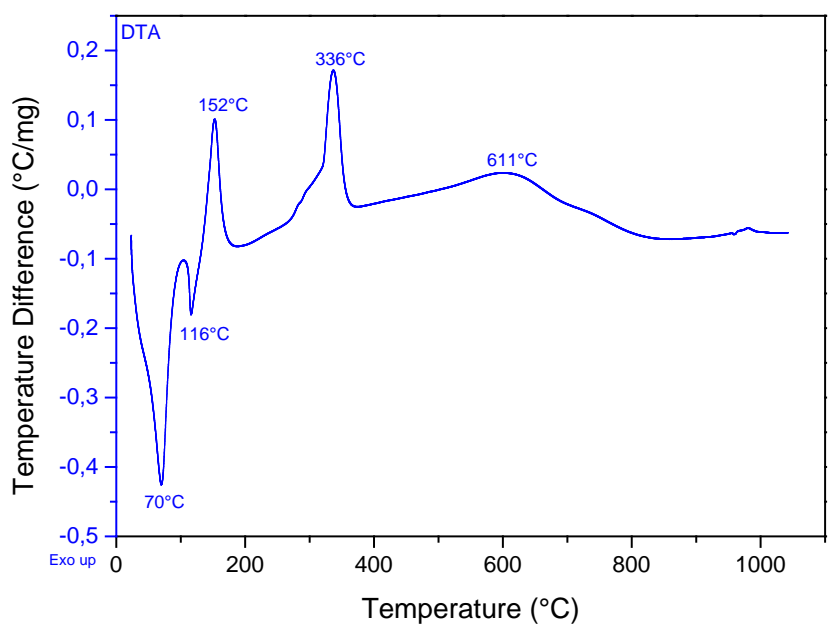

Fig. 12S DTA derivative thermograms of **[Ag(3-cpy)₂(NO₂)]** in air atmosphere.

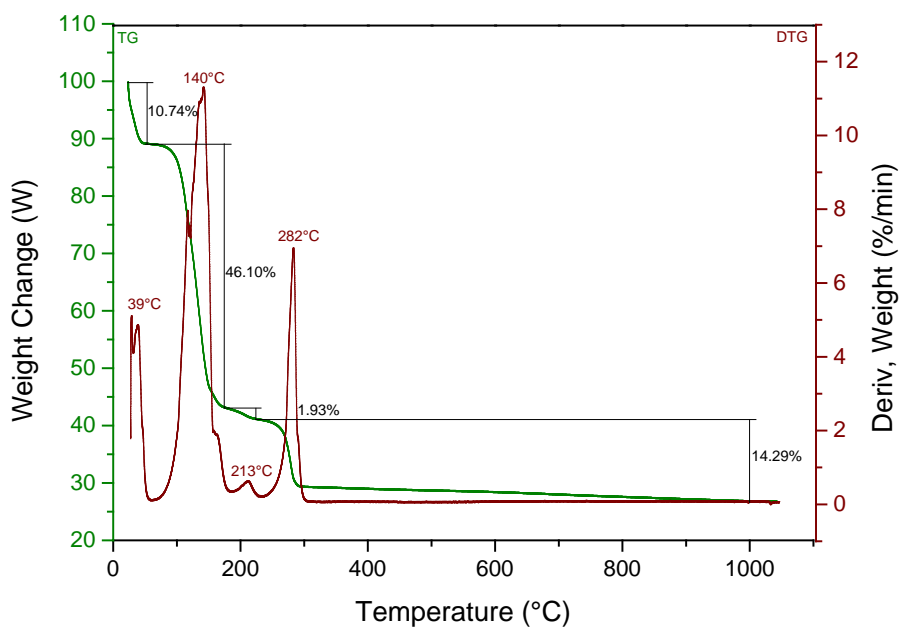

Fig. 13S TG and DTG derivative thermograms of **[Ag(3-cpy)₂(NO₂)]** in N₂ atmosphere.

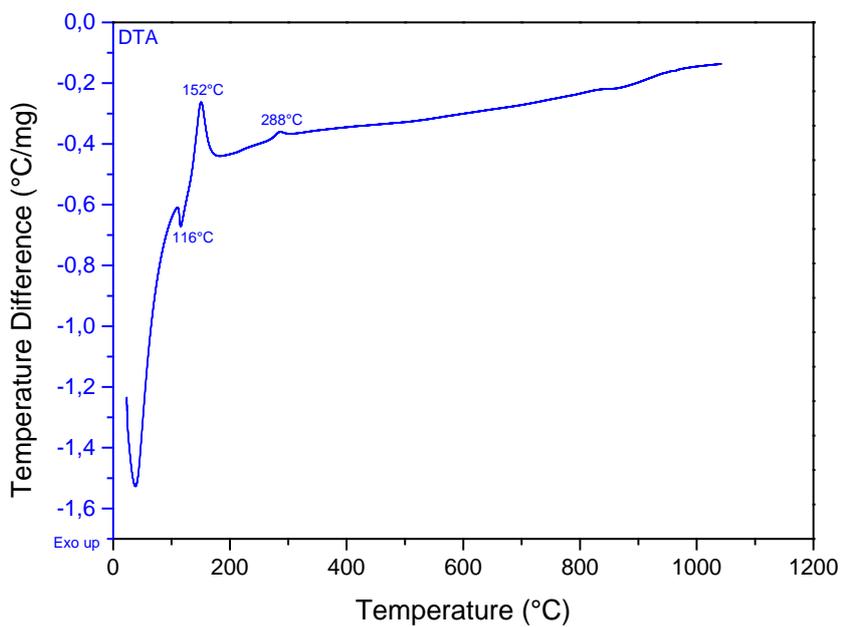

Fig. 14S DTA derivative thermograms of **[Ag(3-cpy)₂(NO₂)]** in N₂ atmosphere.

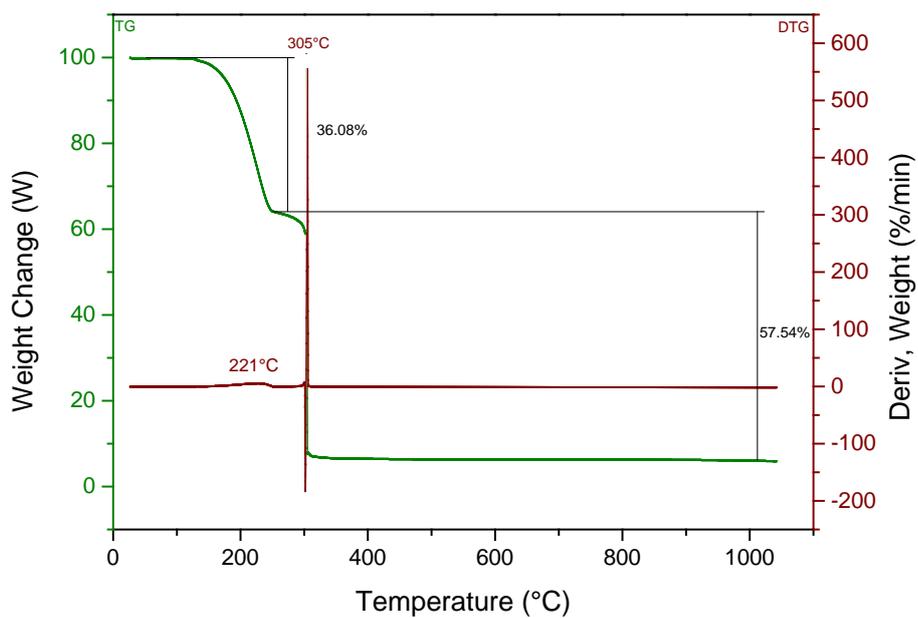

Fig. 15S TG and DTG derivative thermograms of **4** in air atmosphere.

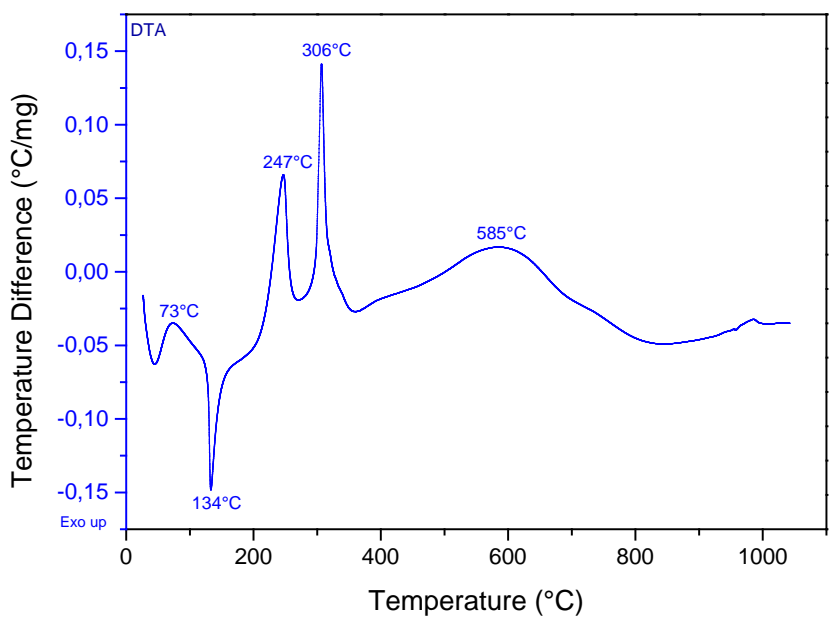

Fig. 16S DTA derivative thermograms of **4** in air atmosphere.

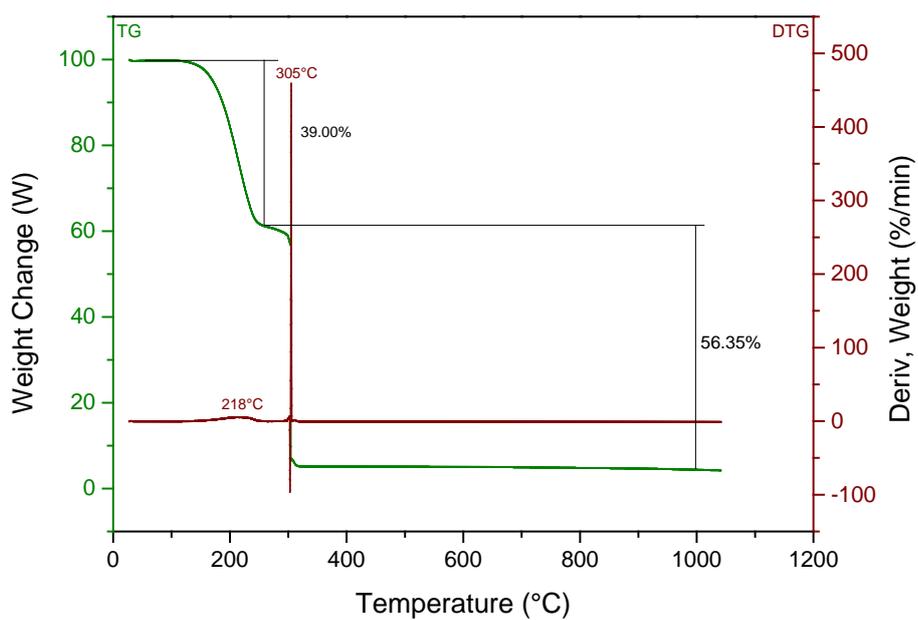

Fig. 17S TG and DTG derivative thermograms of **4** in N$_2$ atmosphere.

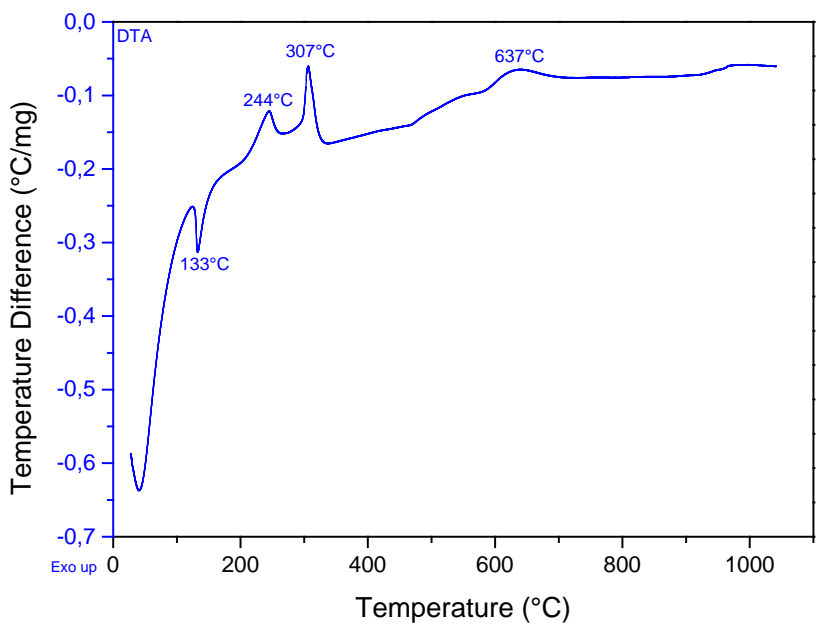

Fig. 18S DTA derivative thermograms of **4** in N$_2$ atmosphere.

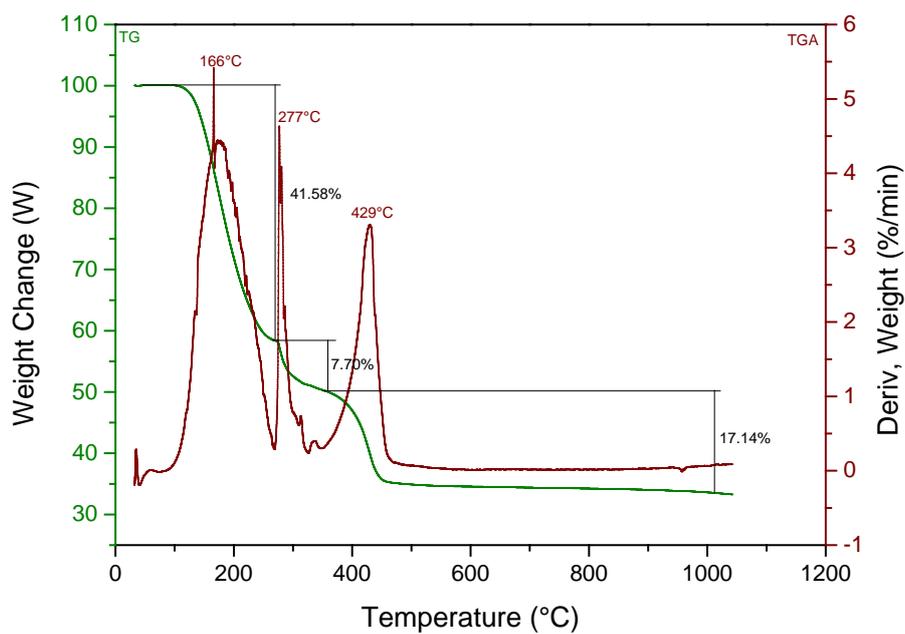

Fig. 19S TG and DTG derivative thermograms of **[Ag(3-cpy)$_2$NO$_3$]$_\infty$** in air atmosphere.

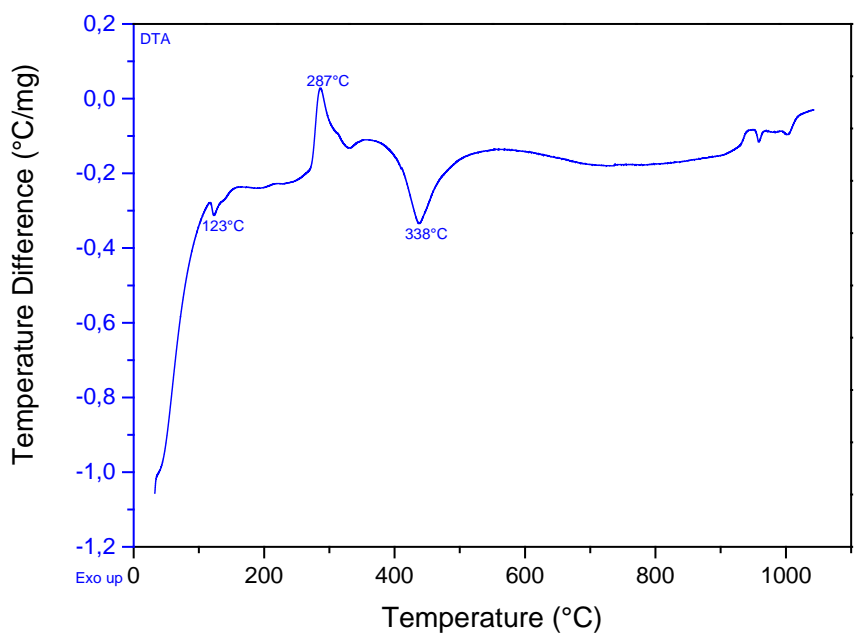

Fig. 20S DTA derivative thermograms of **[Ag(3-cpy)$_2$NO$_3$]$_\infty$** in air atmosphere.

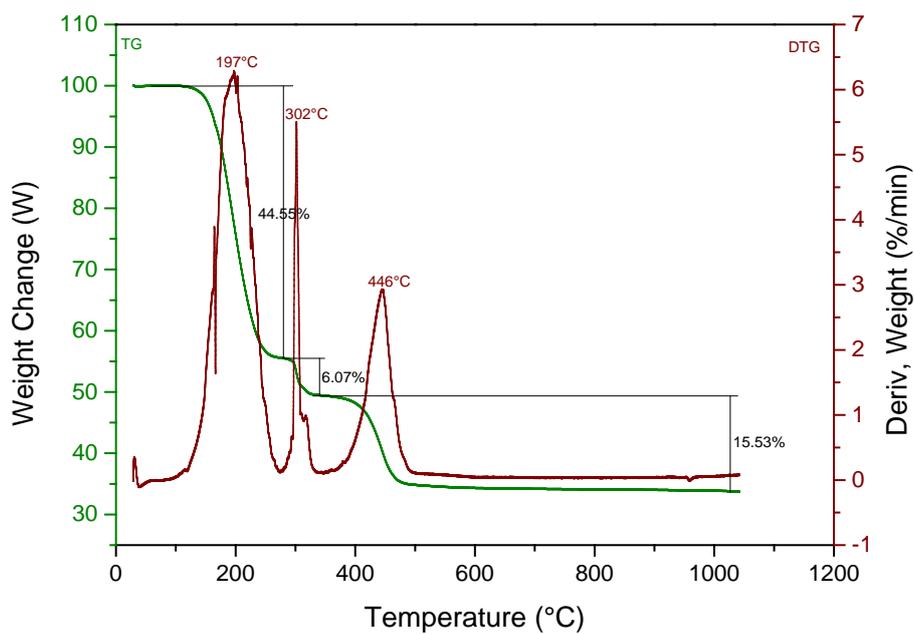

Fig. 21S TG and DTG derivative thermograms of **[Ag₂(1,4-dcb)(NO₃)₂]∞** in air atmosphere.

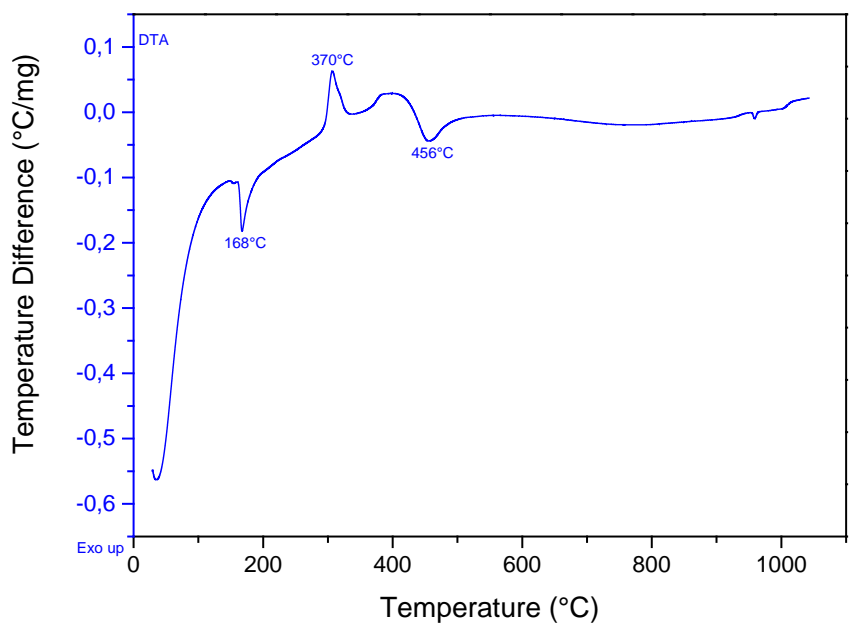

Fig. 22S DTA derivative thermograms of **[Ag₂(1,4-dcb)(NO₃)₂]∞** in air atmosphere.

Table 1S Stepwise thermal degradation data obtained from TG and DTG curves for Ag(I) complexes in air condition.

| Complex | Process | Temperature range[°C] | DTG $T_{max}$[°C] | DTA $T_{max}$[°C] | Weight loss [%] | Ref. |
|---|---|---|---|---|---|---|
| **1** | I | 110-226 | 144 | 153 | 55.55 | |
| | II | 226-327 | 281 | | 10.55 | This work |
| | III | 327-393 | 341 | 344 | 3.92 | |
| **2** | I | 71-152 | 113 | 122 | 38.741 | |
| | II | 152-290 | 249 | 260 | 10.55 | This work |
| | | | | | 14.16 | |
| | III | 299-420 | 384 | 391 | 7.59 | |
| [Ag$_3$(3-cpy)$_2$(NO$_2$)$_3$]$_\infty$ | I | 80-170 | 144 | 155 | 52.87 | |
| | II | 190-287 | 279 | 286 | 3.19 | 17 |
| | III | 293-350 | | 320 | 0.95 | |
| [Ag(3-cpy)$_2$(NO$_2$)] | I | 22-70 | 65 | 70 | 33.95 | |
| | II | 70-173 | 139 | 116 | 30.91 | |
| | III | 173-212 | 210 | 152 | 1.67 | 17 |
| | IV | 212-294 | 278 | 336 | 10,23 | |
| | V | 294-350 | 333 | 611 | 3.22 | |
| **4** | I | 131-249 | 221 | 73 | 36.08 | |
| | II | | | 134 | | |
| | III | 250-310 | | 247 | 57.54 | This work |
| | IV | | 305 | 306 | | |
| | V | | | 585 | | |
| [Ag(3-cpy)$_2$NO$_3$]$_\infty$ | I | 95-270 | 166 | 123 | 41.58 | |
| | II | 270-359 | 277 | 287 | 7.70 | 17 |
| | III | 359-550 | 429 | 338 | 17.14 | |
| [Ag$_2$(1,4-dcb)(NO$_3$)$_2$]$_\infty$ | I | 100-279 | 197 | 168 | 44.55 | |
| | II | 279-341 | 302 | 370 | 6.07 | 17 |
| | III | 341-600 | 446 | 456 | 15.53 | |

Table 2S Stepwise thermal degradation data obtained from TG and DTG curves for Ag(I) complexes in N$_2$ condition.

| Complex | Process | Temperature range[°C] | DTG $T_{max}$[°C] | DTA $T_{max}$[°C] | Weight loss [%] | Ref. |
|---|---|---|---|---|---|---|
| **4** | I | 116-259 | 218 | 133 | 39.00 | |
| | II | | | 244 | | This work |
| | III | 259-303 | 305 | 307 | 56.35 | |
| | IV | | | 637 | | |
| [Ag(3-cpy)$_2$(NO$_2$)] | I | 27-59 | 39 | | 10.74 | |
| | II | 68-175 | 140 | 116 | 46.10 | 17 |
| | III | 180-220 | 213 | 152 | 1.93 | |
| | IV | 250-300 | 282 | 288 | 14.29 | |